\documentclass{sig-alternate-lbnl}

\usepackage{graphics,graphicx}
\graphicspath{{figures/}}
\usepackage[tight,footnotesize]{subfigure}
\usepackage{url,color}
\usepackage{multirow}
\usepackage{clrs+} 

\usepackage{algorithm}
\usepackage[noend]{algpseudocode}

\renewcommand{\algorithmicrequire}{\textbf{Input:}}
\renewcommand{\algorithmicensure}{\textbf{Output:}}
\renewcommand{\algorithmicindent}{12pt}
\usepackage{cite}

\usepackage[colorlinks=true, 
  pdfpagelabels=false,   
	bookmarks=true,        
	bookmarksnumbered=true,
	hypertexnames=false,
	breaklinks=true
	]{hyperref}

\newcommand{\vsb}[0]{\vspace{-6pt}} 

\newcommand{\transpose}     {^{\mbox{\scriptsize \sf T}}}

\newcommand{\lilabel}[1]        {\label{li:#1}}
\newcommand{\liref}[1]      {line~\ref{li:#1}}

\newcommand{\lirefs}[2]     {lines \ref{li:#1}--\ref{li:#2}}

\def\naive{na\"\i ve }

\begin{document}

%
\conferenceinfo{}{}

\title{Parallel Breadth-First Search on\\Distributed Memory Systems}

\numberofauthors{1}
\author{
Ayd{\i}n Bulu\c{c}~~~~~Kamesh Madduri\\
       \affaddr{Computational Research Division}\\
       \affaddr{Lawrence Berkeley National Laboratory}\\
       \affaddr{Berkeley, CA}\\
       \email{\{ABuluc, KMadduri\}@lbl.gov}
}
\toappear{ \rule{\columnwidth}{1pt} This work was supported by the Director, Office of Science, U.S. Department of Energy under Contract No. DE-AC02-05CH11231. This document was prepared as an account of work sponsored by the United States Government. While this document is believed to contain correct information, neither the United States Government nor any agency thereof, nor the Regents of the University of California, nor any of their employees, makes any warranty, express or implied, or assumes any legal responsibility for the accuracy, completeness, or usefulness of any information, apparatus, product, or process disclosed, or represents that its use would not infringe privately owned rights. Reference herein to any specific commercial product, process, or service by its trade name, trademark, manufacturer, or otherwise, does not necessarily constitute or imply its endorsement, recommendation, or favoring by the United States Government or any agency thereof, or the Regents of the University of California. The views and opinions of authors expressed herein do not necessarily state or reflect those of the United States Government or any agency thereof or the Regents of the University of California.}
\maketitle
\begin{abstract}
Data-intensive, graph-based computations are pervasive in several scientific applications, and are known to to be quite challenging to implement on distributed memory systems. In this work, we explore the design space of parallel algorithms for Breadth-First Search (BFS), a key subroutine in several graph algorithms. We present two highly-tuned parallel approaches for BFS on large parallel systems: a level-synchronous strategy that relies on a simple vertex-based partitioning of the graph, and a two-dimensional sparse matrix-partitioning-based approach that mitigates parallel communication overhead. For both approaches, we also present
hybrid versions with intra-node multithreading. Our novel hybrid two-dimensional algorithm reduces communication times by up to a factor of 3.5, relative to a common vertex based approach. Our experimental study identifies execution regimes in which these approaches will be competitive, and we demonstrate extremely high performance on leading distributed-memory parallel systems. For instance, for a 40,000-core parallel execution on Hopper, an AMD Magny-Cours based system, we achieve a BFS performance rate of 17.8 billion edge visits per second on an undirected graph of 4.3 billion vertices and 68.7 billion edges with skewed degree distribution. 

\end{abstract}


\section{Introduction}

The use of graph abstractions to analyze and understand social interaction data, complex engineered systems such as the power grid and the Internet, communication data such as email and phone networks, data from sensor networks, biological systems, and in general, various forms of relational data, has been gaining ever-increasing importance. Common graph-theoretic problems arising in these application areas include identifying and ranking important entities, detecting anomalous patterns or sudden changes in networks, finding tightly interconnected clusters of entities, and so on. The solutions to these problems typically involve classical algorithms for problems such as finding spanning trees, shortest paths, biconnected components, matchings, flow-based computations, in these graphs. To cater to the graph-theoretic analyses demands of emerging ``big data'' applications, it is essential that we speed up the underlying graph problems on current parallel systems.

We study the problem of traversing large graphs in this paper. A traversal refers to a systematic method of exploring all the vertices and edges in a graph. The ordering of vertices using a ``breadth-first'' search (BFS) is of particular interest in many graph problems. Theoretical analysis in the RAM model of computation indicates that the computational work performed by an efficient BFS algorithm would scale linearly with the number of vertices and edges, and there are several well-known serial and parallel BFS algorithms (discussed in Section~\ref{s:prev-work}). However, efficient RAM algorithms do not easily translate into ``good performance'' on current computing platforms. This mismatch arises due to the fact that current architectures lean towards efficient execution of regular computations with low memory footprints, and heavily penalize memory-intensive codes with irregular memory accesses. Graph traversal problems such as BFS are by definition predominantly memory access-bound, and these accesses are further dependent on the structure of the input graph, thereby making the algorithms ``irregular''. 

The recently-created Graph 500 list\footnote{Graph 500, \url{http://www.graph500.org}, last accessed Apr 2011.}, which ranks supercomputers based on their performance on data-intensive applications, chose BFS as their first representative benchmark. Today, distributed memory architectures dominate the supercomputer market and computational scientists have a good understanding of how to map conventional numerical applications to these architectures. By constrast, little is known about the best practices of running a data-intensive graph algorithm and the trade-offs involved. Consequently, the current Graph 500 list is not based on the inherent capabilities of the architectures, but it is based on the quality of various benchmark implementations. 

We present new parallel algorithms and discuss optimized BFS implementations on current distributed-memory systems with multicore processors. BFS on distributed-memory systems involves explicit communication between processors, and the distribution (or partitioning) of the graph among processors also impacts performance. We utilize a testbed of large-scale graphs with billions of vertices and edges to empirically evaluate the performance of our BFS algorithms. These graphs are all sparse, i.e., the number of edges $m$ is just a constant factor times the number of vertices $n$. Further, the average path length in these graphs is a small constant value compared to the number of vertices, or is at most bounded by $\log n$.

\textbf{Our Contributions}
We present two complementary approaches to distributed-memory BFS on graphs with skewed degree distribution. The first approach is a more traditional scheme using one-dimensional distributed adjacency arrays for representing the graph. The second method utilizes a sparse matrix representation of the graph and a two-dimensional partitioning among processors. The following are our major contributions:
\begin{itemize}
\item We present the highest-reported performance numbers (in terms of per-node efficiency using the Graph 500 measure) for BFS on current large-scale distributed memory systems.
\item Our two-dimensional partitioning-based approach, coupled with intranode multithreading, reduces the communication overhead at high process concurrencies by a factor of $3.5$. 
\item Both our approaches include extensive intra-node multicore tuning and performance optimization. The single-node performance of our graph-based approach is comparable to, or exceeds, recent single-node shared memory results on a variety of real-world and synthetic networks. The hybrid schemes enable BFS scalability up to 40,000 cores. 
\item To accurately quantify the memory access costs in BFS, we present a simple memory-reference centric performance model. This model succinctly captures the differences between our two BFS strategies and also provides insight into architectural trends that enable high-performance graph algorithms.
\end{itemize}

{\noindent \bf  Impact on Larger Scale Systems:} Our algorithms address inter-node bandwidth limitations. Therefore, the advantages of our approach 
are likely to grow on future systems since the bisection bandwidth is one of the slowest scaling components in supercomputers. For example, 
the next generation Cray XE6 system (Hopper2) hosted at NERSC has bisection bandwidth comparable to the current generation Cray XT4 (Franklin) 
although Hopper2 has 4 times many cores than Franklin. 
As the cores to bandwidth ratio increases, more and more of the compute capability goes unused with communication-bound algorithms.

\section{Breadth-First Search Overview}
\label{s:prev-work}

\subsection{Preliminaries}

Given a distinguished ``source vertex'' $s$, Breadth-First Search (BFS) systematically explores the graph $G$ to discover every vertex that is reachable from $s$. Let $V$ and $E$ refer to the vertex and edge sets of $G$, whose cardinalities are $n=|V|$ and $m=|E|$. We assume that the graph is unweighted; equivalently, each edge $e \in E$ is assigned a weight of unity. A \emph{path} from vertex $s$ to $t$ is defined as a sequence of edges $\langle u_{i}, u_{i+1} \rangle$ (edge directivity assumed to be $u_i \rightarrow u_{i+1}$ in case of directed graphs), $0 \leq i < l$, where $u_{0} = s$ and $u_{l} = t$. The \emph{length} of a path is the number of hops it travels (or the sum of the weights of edges in the case of weighted graphs). We use $d(s, t)$ to denote the \emph{distance} between vertices $s$ and $t$, or the length of the shortest path connecting $s$ and $t$. BFS implies that all vertices at a distance $k$ (or ``level'' $k$) from vertex $s$ should be first ``visited'' before vertices at distance $k+1$. The distance from $s$ to each reachable vertex is typically the final output. In applications based on a breadth-first graph traversal, one might optionally perform auxiliary computations when visiting a vertex for the first time.  Additionally, a ``breadth-first spanning tree'' rooted at $s$ containing all the reachable vertices can also be maintained. 

\begin{algorithm}[ht]
\begin{algorithmic}[1]
\Require $G(V,E)$, source vertex $s$.
\Ensure $d[1..n]$, where $d[v]$ gives the length of the shortest path from $s$ to $v \in V$.
\ForAll {$v \in V$} \State     $d[v]$ $\leftarrow$ $\infty$ \EndFor
\State  $d[s] \leftarrow 0$, $level \leftarrow 1$, $FS \leftarrow \phi$, $NS \leftarrow \phi$ 
\State  push $s \rightarrow FS$
\While {$FS \neq \phi$}
\For {each $u$ in $FS$}
\For {each neighbor $v$ of $u$}
\If {$d[v] = \infty$}
\State  push $v \rightarrow NS$
\State	$d[v] \leftarrow level$
\EndIf
\EndFor
\EndFor
\State $FS \leftarrow NS$, $NS \leftarrow \phi$, $level \leftarrow level + 1$
\EndWhile
\end{algorithmic}
\caption{Serial BFS algorithm.}
\label{alg:BFS}
\end{algorithm}

Algorithm~\ref{alg:BFS} gives a serial algorithm for BFS. The required breadth-first ordering of vertices is accomplished in this case by using two stacks -- $FS$ and $NS$ -- for storing vertices at the current level (or ``frontier'') and the newly-visited set of vertices (one hop away from the current level) respectively. The number of iterations of the outer while loop (lines 5-11) is bounded by the length of the longest shortest path from $s$ to any reachable vertex $t$. Note that this algorithm is slightly different from the widely-used queue-based serial algorithm~\cite{CLR90}. We can relax the FIFO ordering mandated by a queue at the cost of additional space utilization, but the work complexity in the RAM model is still $O(m+n)$.

\subsection{Parallel BFS: Prior Work}

Parallel algorithms for BFS date back to nearly three decades~\cite{RC78, QD84}. The classical PRAM approach to BFS is a straightforward extension of the serial algorithm presented in Algorithm~\ref{alg:BFS}. The graph traversal loops (lines 6 and 7) are executed in parallel by multiple processing elements, and the distance update and stack push steps (lines 8-10) are atomic. There is a barrier synchronization step once for each level, and thus the execution time in the PRAM model is $O(D)$, where the $D$ is the diameter of the graph. Since the PRAM model does not weigh in synchronization costs, the asymptotic complexity of work performed is identical to the serial algorithm. 

The majority of the novel parallel implementations developed for BFS follow the general structure of this ``level-synchronous'' algorithm, but adapt the algorithm to better fit the underlying parallel architecture. In addition to keeping the parallel work complexity close to $O(m+n)$, the three key optimization directions pursued are
\begin{itemize}
\item ensuring that parallelization of the edge visit steps (lines 6, 7 in Algorithm~\ref{alg:BFS}) is load-balanced,
\item mitigating synchronization costs due to atomic updates and the barrier synchronization at the end of each level, and
\item improving locality of memory references by modifying the graph layout and/or BFS data structures.
\end{itemize}

We discuss recent work on parallel BFS in this section, and categorize them based on the parallel system they were designed for. 


\textbf{Multithreaded systems:} Bader and Madduri~\cite{BM06b} present a fine-grained parallelization of the above level-synchronous algorithm for the Cray MTA-2, a massively multithreaded shared memory parallel system. Their approach utilizes the support for fine-grained, low-overhead synchronization provided on the MTA-2, and ensures that the graph traversal is load-balanced to run on thousands of hardware threads. The MTA-2 system is unique in that it relies completely on hardware multithreading to hide memory latency, as there are no data caches in this system. This feature also eliminates the necessity of tedious locality-improvement optimizations to the BFS algorithm, and Bader and Madduri's implementation achieves a very high system utilization on a 40-processor MTA-2 system. Mizell and Maschhoff~\cite{MM09} discuss an improvement to the Bader-Madduri MTA-2 approach, and present performance results for parallel BFS on a 128-processor Cray XMT system, a successor to the Cray MTA-2. The change specifically is to mitigate thread contention for pushes into the newly visited vertices stack by performing them in batches. The only limit to problem scalability on these uniform shared memory XMT systems is the memory capacity, which can be as large as 100s of gigabytes to a few terabytes. However, overall performance on these systems is bounded by the global memory reference rate that a processor can achieve, which is in turn limited by the network bisection bandwidth. Also, it is difficult to achieve strong parallel scaling for small graphs ($n$ or $m$ comparable to the number of lightweight threads).
 
The current generation of GPGPUs are similar to the Cray XMT systems in their reliance on large-scale multithreading to hide memory latency. In addition, one needs to ensure that the threads perform regular and contiguous memory accesses to achieve high system utilization. This makes optimizing BFS for these architectures quite challenging, as there is no work-efficient way to ensure coalesced accesses to the $d$ array in the level synchronous algorithm. Harish and Narayanan~\cite{HN07} discuss implementations of BFS and other graph algorithms on NVIDIA GPUs. Due to the comparably higher memory bandwidth offered by the GDDR memory, they show that the GPU implementations outperform BFS approaches on the CPU for various low-diameter graph families with tens of millions of vertices and edges. Luo et al.~\cite{LWH10} present an improvement to this work with a new hierarchical data structure to store and access the frontier vertices, and demonstrate that their algorithm is up to $10\times$ faster than the Harish-Narayanan algorithm on recent NVIDIA GPUs and low-diameter sparse graphs. You et al.~\cite{YCY09} study BFS-like traversal optimization on GPUs and multicore CPUs in the context of implementing an inference engine for a speech recognition application. They explore alternatives to atomics in creating the stack of newly-visited vertices. In particular, one could avoid the ``not visited'' check (line 8), aggregate all the edges out of the frontier vertices, sort them by the destination vertex $v$, and then perform the visited check. This ``aggregation-based'' approach is an alternative to atomics, but incurs a much higher computational cost. We will further analyze this variant in the next section.

\textbf{Multicore systems:} There has been a spurt in recent work on BFS approaches for multicore CPU systems. Current x86 multicore architectures, with 8 to 32-way core-level parallelism and 2-4 way simultaneous multithreading, are much more amenable to coarse-grained load balancing in comparison to the multithreaded architectures. Possible $p$-way partitioning of vertices and/or replication of high-contention data structures alleviates some of the synchronization overhead. However, due to the memory-intensive nature of BFS,  performance is still quite dependent on the graph size, as well as the sizes and memory bandwidths of the various levels of the cache hierarchy. Recent work on parallelization of the queue-based algorithm by Agarwal et al.~\cite{APPB10} notes a problem with scaling of atomic intrinsics on multi-socket Intel Nehalem systems. To mitigate this, they suggest a partitioning of vertices and corresponding edges among multiple sockets, and a combination of the fine-grained approach and the accumulation-based approach in edge traversal. In specific, the distance values (or the ``visited'' statuses of vertices in their work) of local vertices are updated atomically, while non-local vertices are held back to avoid coherence traffic due to cache line invalidations. They achieve very good scaling going from one to four sockets with this optimization, at the expense of introducing an additional barrier synchronization for each BFS level. Xia and Prasanna~\cite{XP09} also explore synchronization-reducing optimizations for BFS on Intel Nehalem multicore systems. Their new contribution is a low-overhead ``adaptive barrier'' at the end of each frontier expansion that adjusts the number of threads participating in traversal based on an estimate of work to be performed. They show significant performance improvements over \naive parallel BFS implementations on dual-socket Nehalem systems. Leiserson and Schardl~\cite{LS10} explore a different optimization: they replace the shared queue with a new ``bag'' data structure which is more amenable for code parallelization with the Cilk++ run-time model. They show that their bag-based implementation also scales well on a dual-socket Nehalem system for selected low diameter benchmark graphs. These three approaches use seemingly independent optimizations and different graph families to evaluate performance on, which makes it difficult to do a head-to-head comparison. Since our target architecture in this study are clusters of multicore nodes, we share some similarities to these approaches. We will revisit key aspects of these implementations in the next section, and attempt to distinguish the new contributions of our work.

\textbf{Distributed memory systems:} The general structure of the level-synchronous approach holds in case of distributed memory implementations as well, but fine-grained ``visited'' checks are replaced by edge aggregation-based strategies. With a distributed graph and a distributed $d$ array, a processor cannot tell whether a non-local vertex has been previously visited or not. So the common approach taken is to just accumulate all edges corresponding to non-local vertices, and send them to the owner processor at the end of a local traversal. There is thus an additional all-to-all communication step at the end of each frontier expansion. Interprocessor communication is considered a significant performance bottleneck in prior work on distributed graph algorithms~\cite{CDT05, LGH07}. The relative costs of inter-processor communication and local computation depends on the quality of the graph partitioning and the topological characteristics of the interconnection network. As mentioned earlier, the edge aggregation strategy introduces extraneous computation (which becomes much more pronounced in a fully distributed setting), due to which the level-synchronous algorithm deviates from the $O(m+n)$ work bound. 

The BFS implementation of Scarpazza et al.~\cite{SVP08} for the Cell/B.E. processor, while being a multicore implementation, shares a lot of similarities with the general ``explicit partitioning and edge aggregation'' BFS strategy for distributed memory system. The implementation by Yoo et al.~\cite{YCH05} for the BlueGene/L system is a notable distributed memory parallelization. The authors observe that a two-dimensional graph partitioning scheme would limit key collective communication phases of the algorithms to at most $\sqrt{p}$ processors, thus avoiding the expensive all-to-all communication steps. This enables them to scale BFS to process concurrencies as high as 32,000 processors. However, this implementation assumes that the graph families under exploration would have a regular degree distribution, and computes bounds for inter-process communication message buffers based on this assumption. Such large-scale scalability with or without 2D graph decomposition may not be realizable for graphs with skewed degree distributions. Furthermore, the computation time increases dramatically (up to 10-fold) with increasing processor counts, under a weak scaling regime. This implies that the sequential kernels and data structures used in this study were not work-efficient. As opposed to Yoo et al.'s work, we give details of the data structures and algorithms that are local to each processor in Section~\ref{sec:implementation}. 

Cong et al.~\cite{CAS10} study the design and implementation of several graph algorithms using the partitioned global address space (PGAS) programming model. PGAS languages and runtime systems hide cumbersome details of message passing-based distributed memory implementations behind a shared memory abstraction, while offering the programmer some control over data locality. Cong's work attempts to bridge the gap between PRAM algorithms and PGAS implementations, again with collective communication optimizations. 
Recently, Edmonds et al.~\cite{edmonds10:hipc-srs} gave the first hybrid-parallel 1D BFS implementation that uses active messages.

Software systems for large-scale distributed graph algorithm design include the Parallel Boost graph library~\cite{GL05}, the Pregel~\cite{MAB10} framework. Both these systems adopt a straightforward level-synchronous approach for BFS and related problems. Prior distributed graph algorithms are predominantly designed for ``shared-nothing'' settings. However, current systems offer a significant amount of parallelism within a single processing node, with per-node memory capacities increasing as well. Our paper focuses on graph traversal algorithm design in such a scenario. We present these new parallel strategies and quantify the performance benefits achieved in Section~\ref{s:dist-bfs}. 

\textbf{External memory algorithms:} Random accesses to disk are extremely expensive, and so locality-improvement optimizations are the key focus of external memory graph algorithms. External memory graph algorithms build on known I/O-optimal strategies for sorting and scanning. Ajwani and Meyer~\cite{AM09, ADM06} discuss the state-of-the-art algorithms for BFS and related graph traversal problems, and present performance results on large-scale graphs from several families. Recent work by Pierce et al.~\cite{PGA10} investigates implementations of semi-external BFS, shortest paths, and connected components. 

\textbf{Other Parallel BFS Algorithms:} There are several alternate parallel algorithms to the level-synchronous approach, but we are unaware of any recent, optimized implementations of these algorithms. The fastest-known algorithm (in the PRAM complexity model) for BFS represents the graph as an incidence matrix, and involves repeatedly squaring this matrix, where the element-wise operations are in the min-plus semiring (see~\cite{GM88} for a detailed discussion). This computes the BFS ordering of the vertices in $O(\log n)$ time in the EREW-PRAM model, but requires $O(n^3)$ processors for achieving these bounds. This is perhaps too work-inefficient for traversing large-scale graphs. The level synchronous approach is also clearly inefficient for high-diameter graphs. A PRAM algorithm designed by Ullman and Yannakakis~\cite{UY90}, based on path-limited searches, is a possible alternative on shared-memory systems. However, it is far more complicated than the simple level-synchronous approach, and has not been empirically evaluated. The graph partitioning-based strategies adopted by Ajwani and Meyer~\cite{AM09} in their external memory traversal of high-diameter graphs may possibly lend themselves to efficient in-memory implementations as well. 

\textbf{Other Related Work:} Graph partitioning is intrinsic to distributed memory graph algorithm design, as it helps bound inter-processor communication traffic. One can further relabel vertices based on partitioning or other heuristics~\cite{CM09, CM69}, and this has the effect of improving memory reference locality and thus improve parallel scaling. 

A sparse graph can analogously be viewed as a sparse matrix, and optimization strategies for linear algebra computations similar to BFS, such as sparse matrix-vector multiplication~\cite{WOV09}, may be translated to the realm of graph algorithms to improve BFS performance as well. We will study this aspect in more detail in the next section. 

Recent research shows prospects of viewing graph algorithms as sparse matrix operations~\cite{unifiedstarp, combblas}.
Our work contributes to that area by exploring the use of sparse-matrix sparse-vector multiplication for BFS for the first time.
The formulation of BFS that is common in combinatorial optimization and artificial intelligence search applications~\cite{BBC03,KS05} is different from the focus of this paper.

\section{\texorpdfstring{Breadth-First Search on\\Distributed Memory Systems}{Breadth-First Search on Distributed Memory Systems}}
\label{s:dist-bfs}

In this section, we briefly describe the high-level parallelization strategy employed in our two distributed BFS schemes with accompanying pseudo-code. Section~\ref{sec:implementation} provides more details about the parallel implementation of these algorithms. The algorithms are seemingly straightforward to implement, but eliciting high performance on current systems requires careful data structure choices and low-level performance tuning. Section~\ref{sec:1danal} provides a rigorous analysis of both the parallel implementations.  

\subsection{BFS with 1D Partitioning}

A natural way of distributing the vertices and edges of a graph on a distributed memory system is to let each 
processor own $n/p$ vertices and all the outgoing edges from those vertices. 
We refer to this partitioning of the graph as `1D partitioning', as it translates to the one-dimensional 
decomposition of the incidence matrix corresponding to the graph. 

The general schematic of the level-synchronous parallel BFS algorithm can be modified to work in a distributed scenario with 
1D partitioning as well. Algorithm~\ref{alg:distBFS} gives the pseudo-code for BFS on a cluster of multicore or multithreaded processors. 
The distance array is also distributed among processes. Every process only maintains the status of vertices it owns, and so the 
traversal loop just becomes an edge aggregation phase. We can utilize multithreading within a process to enumerate the adjacencies. 
However, only the owner process of a vertex can identify whether it is newly visited or not. All the adjacencies of the vertices in the current 
frontier need to be sent to their corresponding owner process, which happens in the All-to-all communication step (line 21) of the algorithm. 
Note that the only thread-level synchronization required is due to the barriers. The rest of the steps, buffer packing and unpacking, can be 
performed by the threads in a data-parallel manner. Sections~\ref{sec:implementation} and \ref{sec:1danal} provide more implementation details
and a rigorous analysis of the algorithm respectively. The key aspects to note in this algorithm, in comparison to the serial level-synchronous 
algorithm (Algorithm~\ref{alg:BFS}), is the extraneous computation (and communication) introduced 
due to the distributed graph scenario: creating the message buffers of cumulative size $O(m)$ and the All-to-all communication step.

\begin{algorithm}[ht]
\begin{algorithmic}[1]
\Require $G(V,E)$, source vertex $s$.
\Ensure $d[1..n]$, where $d[v]$ gives the length of the shortest path from $s$ to $v \in V$.
\ForAll {$v \in V$} \State     $d[v]$ $\leftarrow$ $\infty$ \EndFor
\State  $d[s] \leftarrow 0$, $level \leftarrow 1$, $FS \leftarrow \phi$, $NS \leftarrow \phi$ 
\State $op_{s} \leftarrow find\_owner(s)$
\If {$op_{s} = rank$}
\State push $s \rightarrow FS$
\State $d[s] \leftarrow 0$
\EndIf
\For {$0 \leq j < p$}
\State $SendBuf_{j} \leftarrow \phi$ \Comment{$p$ shared message buffers}
\State $RecvBuf_{j} \leftarrow \phi$ \Comment{for MPI communication}
\State $tBuf_{ij} \leftarrow \phi$ \Comment{thread-local stack for thread $i$}
\EndFor

\While {$FS \neq \phi$}
\For {each $u$ in $FS$ \textbf{in parallel}}
\For {each neighbor $v$ of $u$}
\State $p_{v} \leftarrow find\_owner(v)$
\State push $v \rightarrow tBuf_{ip_{v}}$
\EndFor
\EndFor
\State Thread Barrier
\For {$0 \leq j < p$}
\State Merge thread-local $tBuf_{ij}$'s \textbf{in parallel},
\Statex form $SendBuf_{j}$
\EndFor
\State Thread Barrier
\State \textbf{All-to-all collective step} with the master thread:   
\Statex Send data in $SendBuf$, aggregate 
\Statex newly-visited vertices into $RecvBuf$
\State Thread Barrier
\For {each $u$ in $RecvBuf$ \textbf{in parallel}} 
\If {$d[u] = \infty$}
\State $d[u] \leftarrow level$
\State push $u \rightarrow NS_{i}$
\EndIf
\EndFor
\State Thread Barrier
\State $FS \leftarrow \bigcup NS_{i}$ \Comment{thread-parallel}
\State Thread Barrier
\EndWhile
\end{algorithmic}
\caption{Hybrid parallel BFS with vertex partitioning.}
\label{alg:distBFS}
\end{algorithm}


\subsection{BFS with 2D Partitioning}

We next describe a parallel BFS approach that directly works with the sparse adjacency matrix of the graph. 
Factoring out the underlying algebra, each BFS iteration is computationally equivalent to a sparse matrix-sparse vector multiplication 
(SpMSV). Let $A$ denote the adjacency matrix of the graph, represented in a sparse boolean format, $x_k$ denotes the $k$th frontier, 
represented as a sparse vector with integer variables. It is easy to see that the exploration of level $k$ in BFS is algebraically equivalent to 
$x_{k+1} \gets A\transpose \otimes x_k \odot \overline{\bigcup_{i=1}^{x_i}}$ (we will omit the transpose and assume that the input is 
pre-transposed for the rest of this section). The syntax $\otimes$ denotes the matrix-vector 
multiplication operation on a special (select,max)-semiring, $\odot$ denotes element-wise multiplication, and overline represents 
the complement operation. In other words, $\overline{v}_i = 0$ for $v_i \neq 0$ and $\overline{v}_i = 1$ for $v_i = 0$. 
The algorithm does not have to store the previous frontiers explicitly as multiple sparse vectors. In practice, it keeps a
dense $\id{parents} = \bigcup_{i=1}^{x_i}$ array, which is more space efficient and easier to update. 

Our sparse matrix approach uses the alternative 2D decomposition of the adjacency matrix of the graph. Consider the simple checkerboard partitioning,
where processors are logically organized on a square $p = p_r \times p_c$ mesh, indexed by their row and column indices so that the 
$(i,j)$th processor is denoted by $P(i,j)$. Edges and vertices (sub-matrices) are assigned to processors according to a 2D block decomposition. 
Each node stores a sub-matrix of dimensions $(n/p_r) \times (n/p_c)$ in its local memory. For example, $A$ can be partitioned as 
shown below and $A_{ij}$ is assigned to processor $P(i,j)$. 

\begin{equation}
A = \left( 
\begin{array}{c|c|c}
A_{1,1} & \ldots  & A_{1,p_c} \\
\hline
\vdots  & \ddots & \vdots  \\
\hline
A_{p_r,1} & \ldots   & A_{p_r,p_c} 
\end{array} 
\right)
\label{eqn:2dpartitioning}
\end{equation}

\begin{figure}
\centering
\includegraphics[scale=0.58]{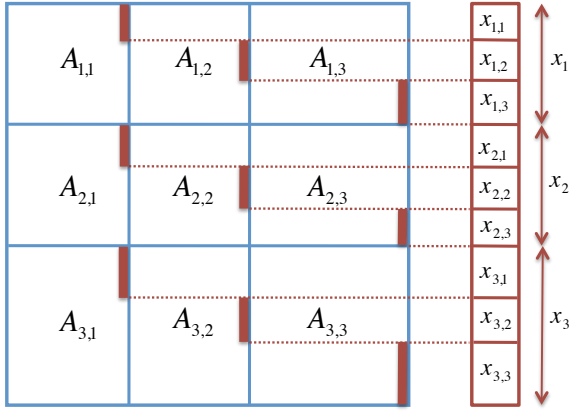}
\caption{
2D vector distribution illustrated by interleaving it with the matrix distribution. 
\label{fig:2ddist}}
\end{figure}

Algorithm~\ref{alg:2dbfs} gives the high-level pseudocode of our parallel algorithm for BFS on 2D-distributed graphs. 
This algorithm implicitly computes the ``breadth-first spanning tree''
by returning a dense parents array. The inner loop block starting in \liref{bfsiter<} performs 
a single level traversal. All vectors and the input matrix are 2D-distributed as illustrated in Figure~\ref{fig:2ddist}.
$f$, which is initially an empty sparse vector, represents the current frontier.
$t$ is an sparse vector that holds the temporary parent information for that iteration only. 
For a distributed vector $v$, the syntax $v_{ij}$ denotes the local $n/p$ sized piece of the vector owned by the $P(i,j)$th 
processor. The syntax $v_i$ denotes the hypothetical $n/p_r$ sized piece of the vector collectively owned by all the processors 
along the $i$th processor row $P(i,:)$. 

Each computational step can be efficiently parallelized with multithreading. The multithreading of the SpMSV operation in \liref{seqspmsv<} 
naturally follows the splitting of the local sparse matrix data structure rowwise to $t$ pieces. The vector operations in 
\lirefs{elementwise<}{parentupdate<} is parallelized simply by exploiting loop-level parallelism. 

\begin{algorithm}[ht]
\begin{algorithmic}[1]
\Require $A$: undirected graph represented by a boolean sparse adjacency matrix, $s$: source vertex id.
\Ensure $\pi$: dense vector, where $\pi[v]$ is the predecessor vertex on the shortest path from $s$ to $v$, 
		or $-1$ if $v$ is unreachable.
\Procedure{BFS\_2D}{A, s}
\State $f(\id{s}) \gets \id{s} $
\For{all processors $P(i,j)$\  \InParallel} 
\While{$f \neq \emptyset$} \lilabel{bfsiter<} 
\State \Call{TransposeVector}{$f_{ij}$}
\State $f_i \gets$ \Call{Allgatherv}{$f_{ij}, P(:,j)$}
\State $ t_i \gets A_{ij} \otimes f_i $  \lilabel{seqspmsv<}
\State $ t_{ij} \gets $ \Call{Alltoallv}{$t_i, P(i,:)$}
\State $ t_{ij} \gets t_{ij} \odot \overline{\pi_{ij}} $ \lilabel{elementwise<}
\State $\pi_{ij} \gets \pi_{ij} + t_{ij} $ \lilabel{parentupdate<}
\State $ f_{ij} \gets t_{ij} $
\EndWhile
\EndFor
\EndProcedure
\end{algorithmic}
\caption{Parallel 2D BFS algorithm.}
\label{alg:2dbfs}
\end{algorithm}

\proc{TransposeVector} redistributes the vector so that the subvector owned by the
$i$th processor row is now owned by the $i$th processor column. In the case of
$p_r=p_c=\sqrt{p}$, the operation is simply a pairwise exchange between $P(i,j)$ and $P(j,i)$. 
In the more general case, it involves an all-to-all exchange among processor groups 
of size $p_r+p_c$. 

\proc{Allgather}($f_{ij}, P(:,j)$) syntax denotes that subvectors $f_{ij}$ for $j=1,..,p_r$ is accumulated
at all the processors on the $j$th processor column. Similarly, \proc{Alltoallv}($t_i, P(i,:)$) denotes that
each processor scatters the corresponding piece of its intermediate vector $t_i$ to its owner, along the $i$th processor row. 

There are two distinct communication phases in a BFS algorithm with 2D partitioning: a pre-computation ``expand'' phase 
over the processor column (with $p_r$ participants), and a 
post-computation ``fold'' phase over the processor row ($p_c$ processes). 
  
In 1D decomposition, the partitioning of frontier vectors naturally follows the vertices. In 2D decomposition, however, vertex
ownerships are more flexible. One practice is to distribute the vector entries only over one processor dimension ($p_r$ or $p_c$) \cite{HendricksonLP95}, 
for instance the diagonal processors if using a square grid ($p_r=p_c$), or the first processors of each processor row. This approach is mostly adequate for sparse matrix-dense vector
multiplication (SpMV), since no local computation is necessary for the ``fold'' phase after the reduce step, to which all the processors 
contribute. For SpMSV, however, distributing the vector to only a subset of processors causes severe imbalance as we show in Section~\ref{sec:distimpl}. 

A more scalable and storage-efficient approach is to let each processor have approximately the same number of vertices. 
In this scheme, which we call the ``2D vector distribution'' in contrast to the ``1D vector distribution'', we still
respect the two-dimensional processor grid. In other words, the 2D vector distribution matches 
the matrix distribution. Each processor row (except the last) is responsible for $t = \lfloor n/p_r \rfloor$ elements. 
The last processor row gets the remaining $ n-\lfloor n/p_r \rfloor (p_r-1)$ elements
Within the processor row, each processor (except the last) is responsible for $l = \lfloor t/p_c \rfloor$ elements. 

In that sense, all the processors on the $i$th processor row contribute to the storage of the vertices that are numbered 
from $v_{ip_r+1}$ to $v_{(i+1)p_r}$. The only downside of the this approach is that the ``expand'' phase becomes an 
all-gather operation (within the processor column) instead of the cheaper broadcast. Another way to view the difference between 
these two approaches is the following. Let $x_i$ be the $i$th subvector of $x$ once we divide 
it equally to $p_r$ pieces. In other words, $x_i = x(ip_r+1:(i+1)p_r)$ spans the vectors 
$v_{ip_r+1}, v_{ip_r+2}, \dots ,v_{(i+1)p_r}$. In both approaches, $x_i$ is owned by some subset of the processors on the $i$th 
processor row. In the first approach it is solely owned by a single $P(i,j)$, whereas in the second approach it is distributed among all 
the processors $P(i,j)$ for $j=1\dots p_c$.

\section{Implementation Details}
\label{sec:implementation}

\subsection{Graph Representation}

For the graph-based BFS implementation, we use a `compressed sparse row' (CSR)-like representation for storing adjacencies. All adjacencies of a vertex are sorted and compactly stored in a contiguous chunk of memory, with adjacencies of vertex $i+1$ next to the adjacencies of $i$. For directed graphs, we store only edges going out of vertices. Each edge (rather the adjacency) is stored twice in case of undirected graphs. An array of size $n+1$ stores the start of each contiguous vertex adjacency block. We use 64-bit integers to represent vertex identifiers. This representation is space-efficient in the sense that the aggregate storage for the distributed data structure is on the same order as the storage that would be needed to store the same data structure serially on a machine with large enough memory. Since our graph is static, linked data structures such as adjacency lists would incur more cache misses without providing any additional benefits.  


On the contrary, a CSR-like representation is too wasteful for storing sub-matrices after 2D partitioning. The aggregate memory required to locally store each submatrix in 
CSR format is $O(n\sqrt{p} + m )$, while storing the whole matrix in CSR format would only take $O(n + m)$. Consequently, a strictly $O(m)$
data structure with fast indexing support is required. The indexing requirement stems from the need to provide near constant time access to individual rows (or columns) 
during the SpMSV operation. One such data structure, doubly-compressed sparse columns (DCSC), has been introduced before~\cite{ipdps08} for hypersparse matrices that 
arise after 2D decomposition. DCSC for BFS consists of an array $\mathsf{IR}$ of row ids (size $m$), which is indexed by two parallel arrays of column pointers ($\mathsf{CP}$)
and column ids ($\mathsf{JC}$). The size of these parallel arrays are on the order of the number of columns that has at least one nonzero ($\id{nzc}$) in them. 

For the hybrid 2D algorihm, we split the node local matrix rowwise to $t$ pieces, as shown in Figure~\ref{fig:nonzerostructure} for two threads. 
Each thread local $n/(p_r t) \times n/p_c$ sparse matrix is stored in DCSC format.

\begin{figure}
\begin{center}
$A =   \left(   
		\begin{array}{c|cccccc}
			& 0 		& 1		& 2		&3 		& 4		& 5  		\\
		\hline
		0	& \times	&    		&  		& \times	& 		&   		\\
		1	& \times 	&  \times   	&  		& 		& 		&   		\\
		2	&  		&     		& 	 	& \times	& 		& \times  	\\
		\hline
		3	& \times 	&     		&  \times	& \times	& 		&   		\\
		4	&  		&  \times   	&  		& 		& 		&   		\\
		5	&  		&     		& \times 	& 		& 		&   			
	        \end{array}\right) $
\end{center}
\caption{Nonzero structure of node-local matrix $A$
\label{fig:nonzerostructure}}
\end{figure}

A compact representation of the frontier vector is also important. It should be represented in a sparse format, where only the indices of the non-zeros are stored. We use a stack in the 1D implementation and a sorted sparse vector in the 2D implementation. Any extra data that are piggybacked to the frontier vectors adversely affect the performance, since the communication volume of the BFS benchmark is directly 
proportional to the size of this vector. 

\subsection{Local Computation}

There are two potential bottlenecks to multithreaded parallel scaling in Algorithm~\ref{alg:distBFS} on our target architectural platforms (multicore systems with modest levels of thread-level parallelism). Consider pushes of newly-visited vertices to the stack $NS$. A shared stack for all threads would involve thread contention for every insertion. An alternative would be to use thread-local stacks (indicated as $NS_{i}$ in the algorithm) for storing these vertices, and merging them at the end of each iteration to form $FS$, the frontier stack. Note that the total number of queue pushes is bounded by $n$, the number of vertices in the graph. Hence, the cumulative memory requirement for these stacks is bounded as well, and the additional computation performed due to merging would be $O(n)$. Our choice is different from the approaches taken in prior work (such as specialized set data structures~\cite{LS10} or a shared queue with atomic increments~\cite{APPB10}). For multithreaded experiments conducted in this study, we found that our choice does not limit performance, and the copying step constitutes a very minor overhead.

Next, consider the distance checks (lines 24-25) and updates in Algorithm~\ref{alg:distBFS}. This is typically made atomic to ensure that a new vertex is added only once to the stack $NS$. However, the BFS algorithm is still correct even if a vertex is added multiple times, as the distance value is guaranteed to be written correctly after the thread barrier and memory fence at the end of a level of exploration. Cache coherence further ensures that the correct value may propagate to other cores once it is updated. We observe that we actually perform a very small percentage of additional insertions (less than $0.5$\%) for all the graphs we experimented with at six-way threading. This lets us avert the issue of non-scaling atomics across multi-socket configurations~\cite{APPB10}. This optimization was also considered by Leiserson et al.~\cite{LS10} (termed ``benign races'') for insertions to their bag data structure.

For the 2D algorithm, the computation time is dominated by the sequential SpMSV operation in \liref{seqspmsv<}  of Algorithm~\ref{alg:2dbfs}. This corresponds to selection, scaling and finally merging columns of the local adjacency matrix that are indexed by the nonzeros in the sparse vector. Computationally, we form the union $\bigcup{A_{ij}(:,k)}$ for all $k$ where $f_i(k)$ exists.

\begin{figure}
\centering
\includegraphics[scale=0.5]{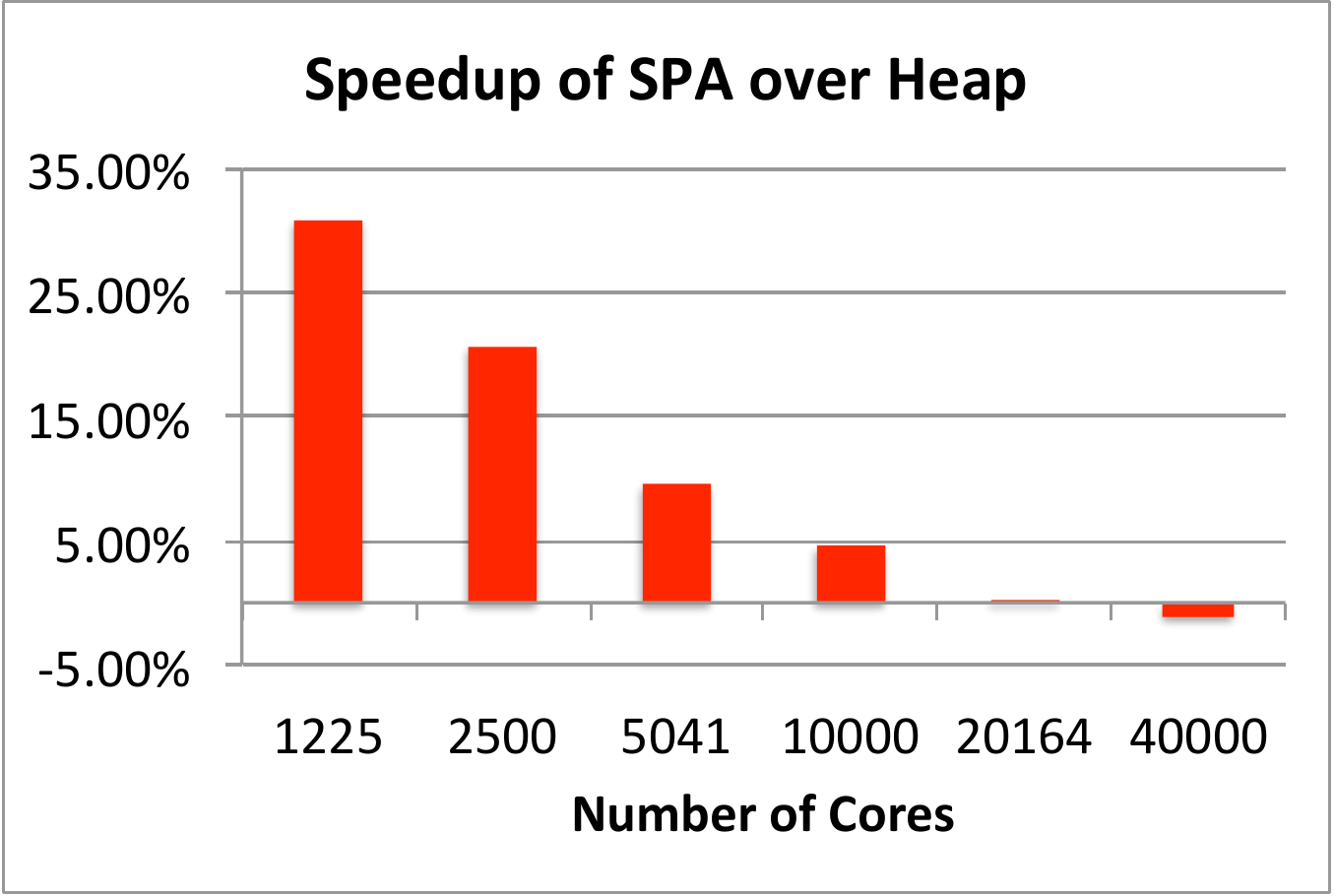}
\caption{
The speedup of using the SPA for local SpMSV operation over using a heap (priority queue) data structure. 
We see that after 10K processors, the difference becomes marginal and heap option becomes preferable due to its lower memory consumption. The experiment is run on Hopper.  
\label{fig:spavsheap}}
\end{figure}

We explored multiple methods of forming this union. The first option is to use a priority-queue of size $\id{nnz}(f_i)$ and perform a unbalanced multiway merging. While this option has the 
advantage of being memory-efficient and automatically creating a sorted output, the extra logarithmic factor hurts the performance at small concurrencies, even after using a highly optimized 
cache-efficient heap. The cumulative requirement for these heaps are $O(m)$. The second option is to use a sparse accumulator (SPA)~\cite{smatlab} which is composed of a dense vector of 
values, a bit mask representing the ``occupied'' flags, and a list that keeps the indices of existing elements in the output vector. The SPA approach proved to be faster for lower concurrencies, 
although it has disadvantages such as increasing the memory footprint due to the temporary dense vectors, and having to explicitly sort the indices at the end of the 
iteration. The cumulative requirement for the sparse accumulators are $O(n \, p_c)$. Figure~\ref{fig:spavsheap} shows the results of a microbenchmark that revealed a transition point around 
10000 cores (for flat MPI version) after which the priority-queue approach is more efficient, both in terms of speed and memory footprint. For a run on 10000 cores, the per core memory 
footprint of the SPA approach on a $100\times 100$ processor grid, running a scale 33 graph, is $9 \cdot 2^{33} /100$ bytes, over 750\,MB. Our final algorithm is therefore a polyalgorithm depending on the concurrency. 

\subsection{Distributed-memory parallelism}
\label{sec:distimpl}

We use the MPI message-passing library to express the inter-node communication steps. In particular, we extensively utilize the collective communication routines Alltoallv, Allreduce, and Allgatherv.

Our 2D implementation relies on the linear-algebraic primitives of the Combinatorial BLAS framework~\cite{combblas}, with certain BFS specific optimizations enabled.

\begin{figure}
\centering
\includegraphics[scale=0.45]{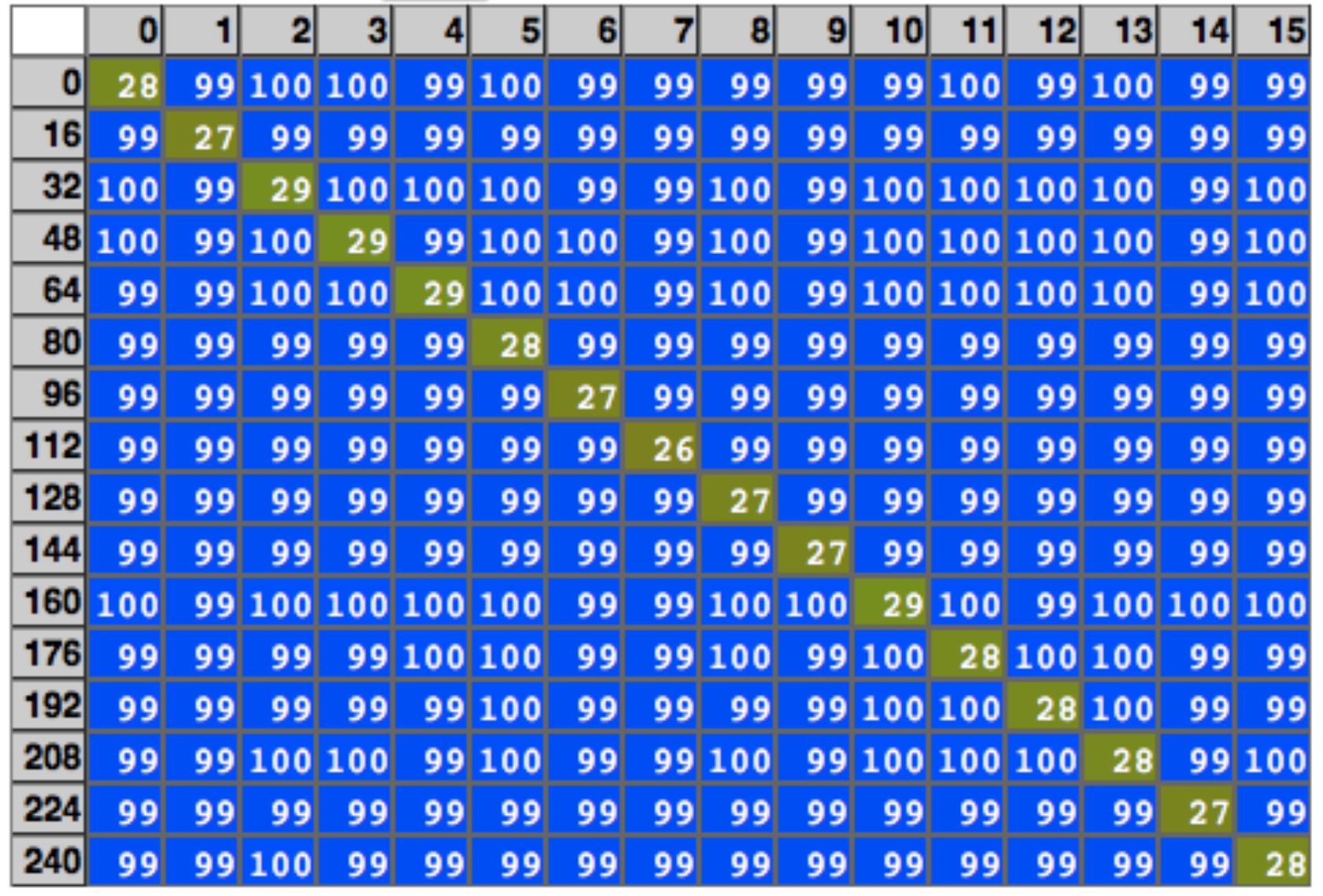}
\caption{
The time spent in MPI calls for a loop that performs BFS iterations followed by a globally synchronizing Allreduce. 
The sparse $x$ and $y$ vectors are distributed to diagonal processors only. 
The numbers are in percentages, normalized to the maximum across all processors. 
The idling times of the waiting processors account for the higher MPI time spent on off-diagonal processors. 
Experiment is performed on 256 processors logically forming a $16\times16$ processor grid. 
\label{fig:mpitime}}
\end{figure}

We chose to  distribute the vectors over all processors instead of only a subset (the diagonal processors in this case) of processors. In SpMSV, the accumulation of sparse contributions require the diagonal processor to go through an additional local merging phase, during which all other processors on the processor row sit idle. The severe load imbalance that results is shown in Figure~\ref{fig:mpitime}, which shows the time spent in MPI calls (normalized to 100\% across all processors). In this experiment, the SpMSV calls are immediately followed by a global Allreduce. The waiting time for this blocking collective is accounted for the total MPI time. We see that the time spent idling is approximately 3-4 times of the time spent in communication. 

Distributing the vectors over all processors (2D vector distribution) remedies this problem and we observe almost no load imbalance in that case.


\subsection{Load-balancing traversal}

We achieve a reasonable load-balanced graph traversal by randomly shuffling all the vertex identifiers prior to partitioning. This leads to each process getting roughly the same number of vertices and edges, regardless of the degree distribution. An identical strategy is also employed in the Graph 500 BFS benchmark. The downside is that the 
edge cut is potentially as high as an average random balanced cut, which can be $O(m)$ for several random graph families~\cite{schreiber:231}. 

\section{Algorithm Analysis}
\label{sec:anal}

Note that the RAM and PRAM models, capturing asymptotic work and parallel execution time complexity with extremely simple underlying assumptions of the machine model, are inadequate to analyze and contrast parallel BFS algorithmic variants on current parallel systems. For instance, the PRAM asymptotic time complexity for a level-synchronous parallel BFS is $O(D)$ (where $D$ is the diameter of the graph), and the work complexity is $O(m+n)$. These terms do not provide any realistic estimate of performance on current parallel systems. 

We propose a simple linear model to capture the cost of regular (unit stride or fixed-stride) and irregular memory references to various levels of the memory hierarchy, as well as to succinctly express inter-processor MPI communication costs. We use the terms $\alpha$ and $\beta$ to account for the latency of memory accesses and the transfer time per memory word (i.e., inverse of bandwidth) respectively. Further, we use $\alpha_{L}$ to indicate memory references to local memory, and $\alpha_{N}$ to denote message latency over the network (remote memory accesses). The bandwidth terms can also be similarly defined. To account for the differing access times to various levels of the memory hierarchy, we additionally qualify the $\alpha$ term to indicate the size of the data structure (in memory words) that is being accessed. $\alpha_{L,x}$, for instance, would indicate the latency of memory access to a memory word in a logically-contiguous memory chunk of size $x$ words. Similarly, to differentiate between various inter-node collective patterns and algorithms, we qualify the network bandwidth terms with the communication pattern. For instance, $\beta_{N, p2p}$ would indicate the sustained memory bandwidth for point-to-point communication, $\beta_{N, a2a}$ would indicate the sustained memory bandwidth per node in case of an all-to-all communication scenario, and $\beta_{N, ag}$ would indicate the sustained memory bandwidth per node for an allgather operation. 

Using synthetic benchmarks, the values of $\alpha$ and $\beta$ defined above can be calculated offline for a particular parallel system and software configuration. The programming model employed, the messaging implementation used, the compiler optimizations employed are some software factors that determine the various $\alpha$ and $\beta$ values.

\subsection{Analysis of the 1D Algorithm}
\label{sec:1danal}

Consider the locality characteristics of memory references in the level-synchronous BFS algorithm. Memory traffic comprises touching every edge once (i.e., accesses to the adjacency arrays, cumulatively $m$), reads and writes from/to the frontier stacks ($n$), distance array checks ($m$ irregular accesses to an array of size $n$) and writes ($n$ accesses to $d$). The complexity of the 1D BFS algorithm in our model is thus $m\beta_{L}$ (cumulative adjacency accesses) + $n\alpha_{n}$ (accesses to adjacency array pointers) + $m\alpha_{n}$ (distance checks/writes). 

In the parallel case with 1D vertex and edge partitioning, the number of local vertices $n_{loc}$ is approximately $n/p$ and the number of local edges is $m/p$. The local memory reference cost is given by $\frac{m}{p}\beta_{L} + \frac{n}{p}\alpha_{L,n/p} + \frac{m}{p}\alpha_{L,n/p}$. The distance array checks thus constitute the substantial fraction of the execution time, since the $\alpha_{L,n/p}$ term is significantly higher than the $\beta_{L}$ term. One benefit of the distributed approach is the array size for random accesses reduces from $n$ to $n/p$, and so the cache working set of the algorithm is substantially lower. Multithreading within a node (say, $t$-way threading) has the effect of reducing the number of processes and increasing the increasing the process-local vertex count by a factor of $t$. 

The remote memory access costs are given by the All-to-all step, which involves a cumulative data volume of $m (p-1)/p$ words sent on the network. For a random graph with a uniform degree distribution, each process would send every other process roughly $m/p^2$ words. This value is typically large enough that the bandwidth component dominates over the latency term. Since we randomly shuffle the vertex identifiers prior to execution of BFS, these communication bounds hold true in case of the synthetic random networks we experimented with in this paper. Thus, the per-node communication cost is given by $p\alpha_{N} + \frac{m}{p}\beta_{N, a2a}(p)$. $\beta_{N, a2a}(p)$ is a function of the processor count, and several factors, including the interconnection topology, node injection bandwidth, the routing protocol, network contention, etc. determine the sustained per-node bandwidth. For instance, if nodes are connected in a 3D torus, it can be shown that bisection bandwidth scales as $p^{2/3}$. Assuming all-to-all communication scales similarly, the communication cost can be revised to $p\alpha_{N} + \frac{m}{p}p^{1/3}\beta_{N, a2a}$. If processors are connected via a ring, then $p\alpha_{N} + \frac{m}{p}p\beta_{N, a2a}$ would be an estimate for the all-to-all communication cost, essentially meaning no parallel speedup.




\subsection{Analysis of the 2D Algorithm}
\label{sec:2danal}


Consider the general 2D case processor grid of $p_r \times p_c$. The size of the local adjacency matrix is $n/p_r \times n/p_c$. The number of memory
references is the same as the 1D case, cumulatively over all processors. However, the cache working set is bigger, because 
the sizes of the local input (frontier) and output vectors are $n/p_r$ and $n/p_c$, respectively. 
The local memory reference cost is given by $\frac{m}{p}\beta_{L} + \frac{n}{p}\alpha_{L,n/p_c} + \frac{m}{p}\alpha_{L,n/p_r}$. 
The higher number of cache misses associated with larger working sets is perhaps the primary reason for the relatively higher 
computation costs of the 2D algorithm.  

Most of the costs due to remote memory accesses is concentrated in two operations. The expand phase is characterized by an Allgatherv 
operation over the processor column (of size $p_r$) and the fold phase is characterized by an
Alltoallv operation over the processor row (of size $p_c$). 

The aggregate input to the Allgatherv step is $O(n)$ over all iterations. 
However, each processor receives a $1/p_c$ portion of it, meaning that frontier subvector gets replicated along the processor column.
Hence, the per node communication cost is $p_r\alpha_{N} + \frac{n}{p_c}\beta_{N, ag}(p_r)$.  
This replication can be partially avoided by performing an extra round of communication where each processor individually examines its 
columns and broadcasts the indices of its nonempty columns. However, this extra step does not decrease the asymptotic complexity for 
RMAT graphs, neither did it gave any performance increase in our experiments.   

The aggregate input to the Alltoallv step can be as high as $O(m)$, although the number is
lower in practice due to in-node aggregation of newly discovered vertices that takes
place before the communication. Since each processor receives only a $1/p$ portion of this data,
the remote costs due to this step are at most $p_c\alpha_{N} + \frac{m}{p}\beta_{N, a2a}(p_c)$.

We see that for large $p$, the expand phase is likely to be more expensive than the fold phase. In fact,
Table~\ref{tab:commdecompose} experimentally confirms the findings of our analysis. We see that Allgatherv always consumes a higher 
percentage of the BFS time than the Alltoallv operation, with the gap widening as the matrix gets sparser. This is because 
for fixed number of edges, increased sparsity only affects the Allgatherv performance, to a first degree approximation. In practice, it slightly
slows down the Alltoallv performance as well, because the in-node aggregation is less effective for sparser graphs. 

\begin{table}
\caption{Decomposition of communication times for the flat (MPI only) 2D algorithm on Franklin, using the R-MAT graphs. 
Allgatherv takes place during the expand phase and Alltoallv takes place during the fold phase. The edge counts are
kept constant.}
\begin{center}
\scalebox{0.9}{
\begin{tabular}{c|c|c|c|c|c}
Core 				& Problem  &  Edge  	& BFS time 	& Allgatherv   	& Alltoallv  	\\ 
count				& scale    &  factor	& (secs)	& (percent.) 	& (percent.)	\\ 
	\hline
\multirow{3}{*}{1024}		& 27 		& 64 		& 2.67  &  7.0\%	& 6.8\%	\\
				& 29		& 16 		& 4.39	&  11.5\%	& 7.7\%	\\
				& 31 		& 4	  	& 7.18  & 16.6\%	& 9.1\%	\\
\hline
\multirow{3}{*}{2025}		& 27 		& 64		& 1.56	&  10.4\%	& 7.6\%	\\
				& 29		& 16	 	& 2.87 	& 19.4\%	& 9.2\%	\\
				& 31 		& 4	 	& 4.41 	& 24.3\%	& 9.0\%	\\
\hline
\multirow{3}{*}{4096}		& 27 		& 64		& 1.31  & 13.1\%	& 7.8\%	\\
				& 29		& 16	 	& 2.23  & 20.8\%	& 9.0\%	\\
				& 31 		& 4	 	& 3.15  & 30.9\%	& 7.7\%	\\
\end{tabular}}
\label{tab:commdecompose}
\end{center}
\end{table}

Our analysis successfully captures that the relatively lower communication costs of the 2D algorithm by
representing $\beta_{N, x}$ as a function of the processor count.


\section{Experimental Studies}
We provide an apples-to-apples comparison of our four different BFS implementations. For both 1D and 2D distributions, we experimented with flat MPI as well as hybrid MPI+Threading versions. We use synthetic graphs based on the R-MAT random graph model~\cite{CZF04}, as well as a big real world graph that represents a web crawl of the UK domain~\cite{BV04} (uk-union). The R-MAT generator creates networks with skewed degree distributions and a very low graph diameter. We set the R-MAT parameters $a$, $b$, $c$, and $d$ to $0.59,0.19,0.19,0.05$ respectively. These parameters are identical to the ones used for generating synthetic instances in the Graph 500 BFS benchmark. R-MAT graphs make for interesting test instances: traversal load-balancing is non-trivial due to the skewed degree distribution, the graphs lack good separators, and common vertex relabeling strategies are also expected to have a minimal effect on cache performance. The diameter of the uk-union graph is significantly higher ($\approx 140$) than R-MAT's (less than $10$), allowing us to access the sensitivity of
our algorithms with respect to the number of synchronizations. We use undirected graphs for all our experiments, but the BFS approaches can work with directed graphs as well. 

To compare performance across multiple systems using a rate analogous to the commonly-used floating point operations/second, we normalize the serial and parallel execution times by the number of edges visited in a BFS traversal and present a `Traversed Edges Per Second' (TEPS) rate. For a graph with a single connected component (or one strongly connected component in case of directed networks), the baseline BFS algorithm would visit every edge twice (once in case of directed graphs). We only consider traversal execution times from vertices that appear in the large component, compute the average time using at least 16 randomly-chosen sources vertices for each benchmark graph, and normalize the time by the cumulative number of edges visited to get the TEPS rate. As suggested by the Graph 500 benchmark, we first symmetrize the input to model undirected graphs. For TEPS calculation, we only count the number of edges in the original directed graph, despite visiting symmetric edges as well. For R-MAT graphs, the default edge count to vertex ratio is set to 16 (which again corresponds to the Graph 500 default setting), but we also vary the ratio of edge to vertex counts in some of our experiments.

We collect performance results on `Franklin', a 9660-node Cray XT4 and `Hopper', a 6392-node Cray XE6. Both supercomputers are located at NERSC, Lawrence Berkeley National Laboratory.
Each XT4 node contains a quad-core 2.3~GHz AMD Opteron processor, which is tightly integrated to the XT4 interconnect via a Cray SeaStar2 ASIC  through a  HyperTransport (HT) 2 interface 
capable of 6.4~GB/s. The SeaStar routing chips are interconnected in a 3D torus topology, and each link is capable of 7.6~GB/s peak bidirectional bandwidth. The 3D torus topology implies that each node has a direct link to its six nearest neighbors. Typical MPI latencies will range from {4.5 - 8.5} $\mu$s, depending on the size of the system and the job placement. The Opteron Budapest 
processor is a superscalar out-of-order core that may complete both a single instruction-multiple data (SIMD) floating-point add and a SIMD floating-point multiply per cycle, the peak 
double-precision floating-point performance (assuming balance between adds and multiplies) is 36.8 GFlop/s. Each core has both a private 64~KB L1 data cache and a 512~KB L2 victim cache. 
The four cores on a socket share a 2~MB L3 cache. The Opteron integrates the memory controllers on-chip and provides an inter-socket network (via HT) to provide cache coherency as well as 
direct access to remote memory. This machine uses DDR2-800 DIMMs providing a DRAM pin bandwidth of 12.8~GB/s.

Each XE6 node contains two twelve-core 2.1~GHz AMD Opteron processors, integrated to the Cray Gemini interconnect through HT~3 interfaces. Each Gemini chip is capable of 
9.8~GB/s bandwidth. Two Hopper nodes share a Gemini chip as opposed to the one-to-one relationship on Hopper. The effective bisection bandwidth of Hopper is slightly 
(ranging from {1-20\%}, depending on the bisection dimension) lower than Franklin's. 
Each twelve-core `MagnyCours' die is essentially composed of two 6-core NUMA nodes. Consequently, we apply 6-way multithreaded for our hybrid
codes in order to decouple the NUMA effects. 

We used the \texttt{gcc 4.5} for compiling both codes. The 1D codes are implemented in C, whereas the 2D codes are implemented in C++. We use Cray's
MPI implementation, which is based on MPICH2. For intra-node threading, we use the GNU C compiler's OpenMP library. For all 2D experiments,
we used the closest square processor grid. For hybrid codes, we applied 4-way multithreading on Franklin and 6-way multithreading on Hopper. 

\begin{figure*}
\centering
\subfigure[$n=2^{29}$, $m=2^{33}$]{\includegraphics[width=0.47\textwidth]{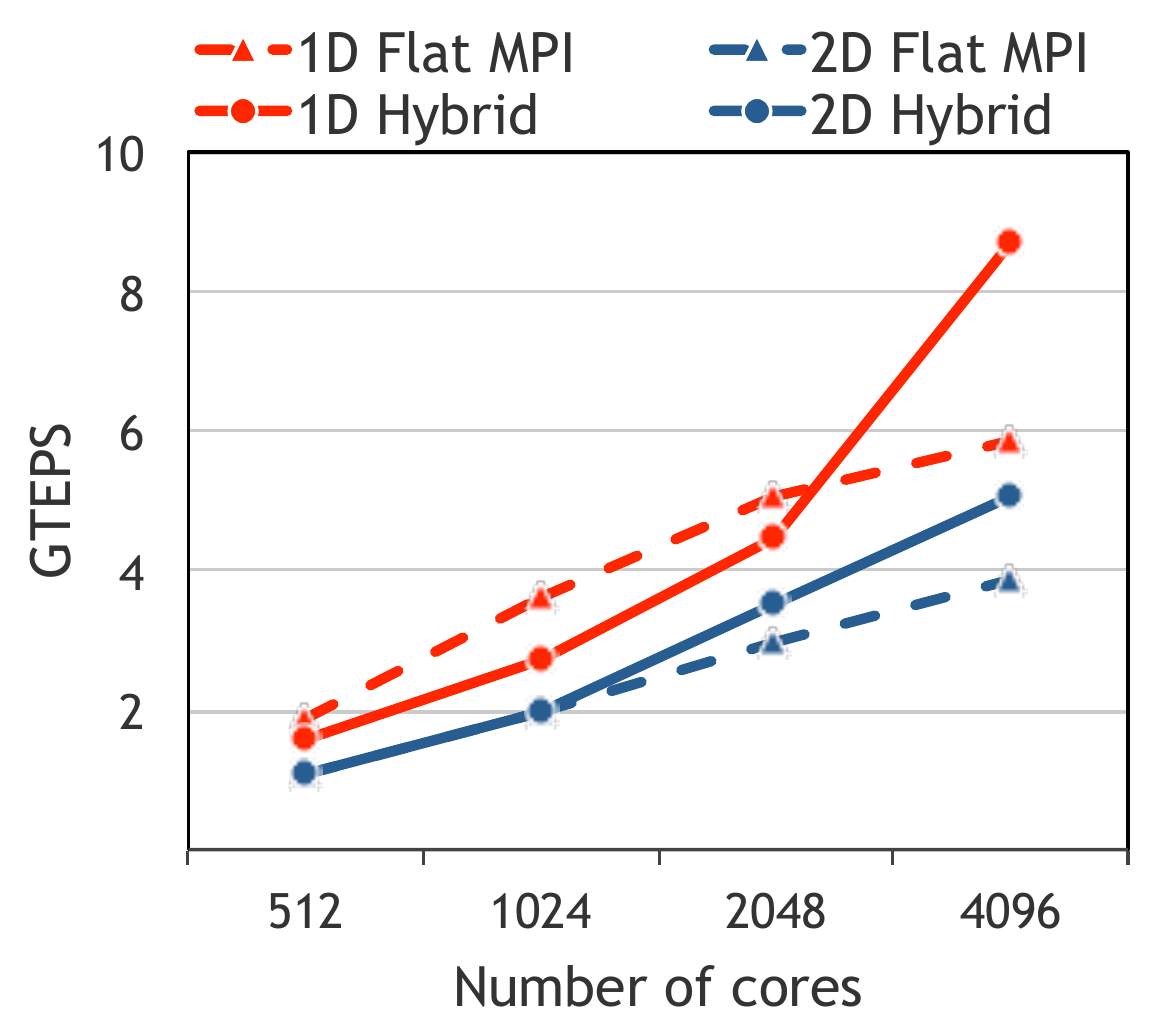}\label{fig:franklin-teps-scale29}}
\subfigure[$n=2^{32}$, $m=2^{36}$]{\includegraphics[width=0.47\textwidth]{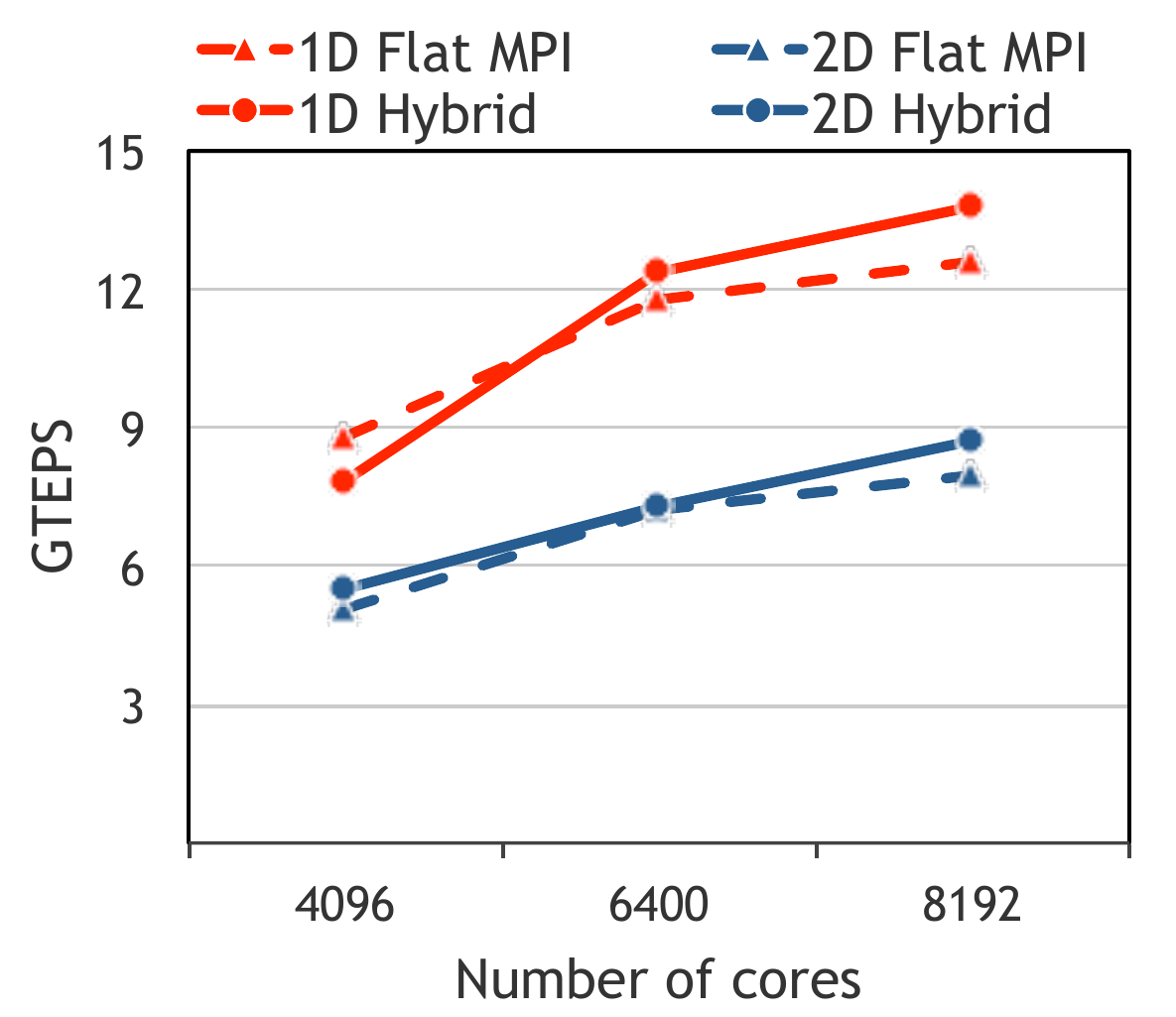}\label{fig:franklin-teps-scale32}}
\caption{BFS `strong scaling' results on Franklin for Graph 500 R-MAT graphs: Performance rate achieved (in GTEPS) on increasing the number of processors.}
\label{fig:franklin-strongscaling}
\end{figure*}

\begin{figure*}
\centering
\subfigure[$n=2^{29}$, $m=2^{33}$]{\includegraphics[width=0.47\textwidth]{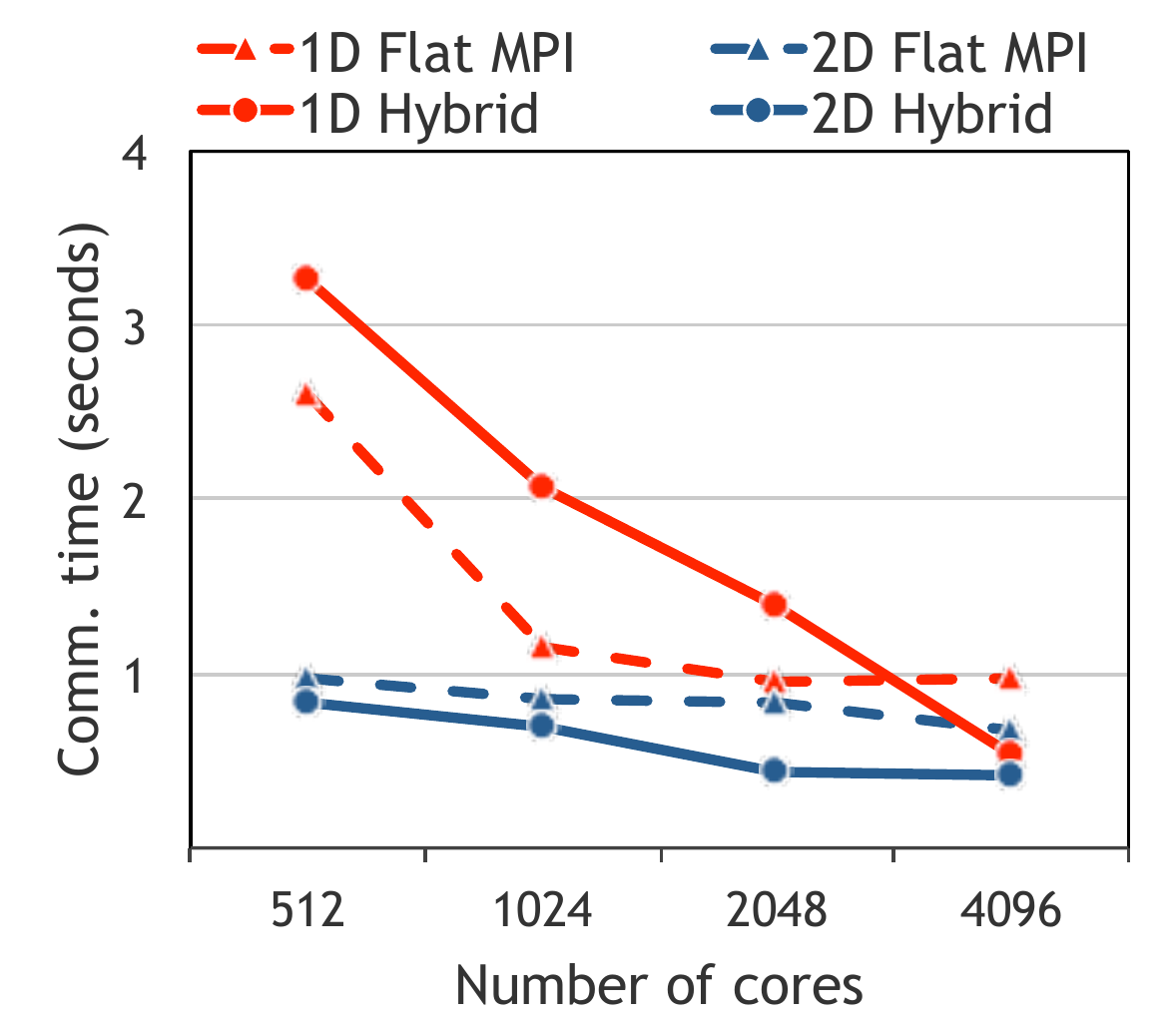}\label{fig:franklin-comm-scale29}}
\subfigure[$n=2^{32}$, $m=2^{36}$]{\includegraphics[width=0.46\textwidth]{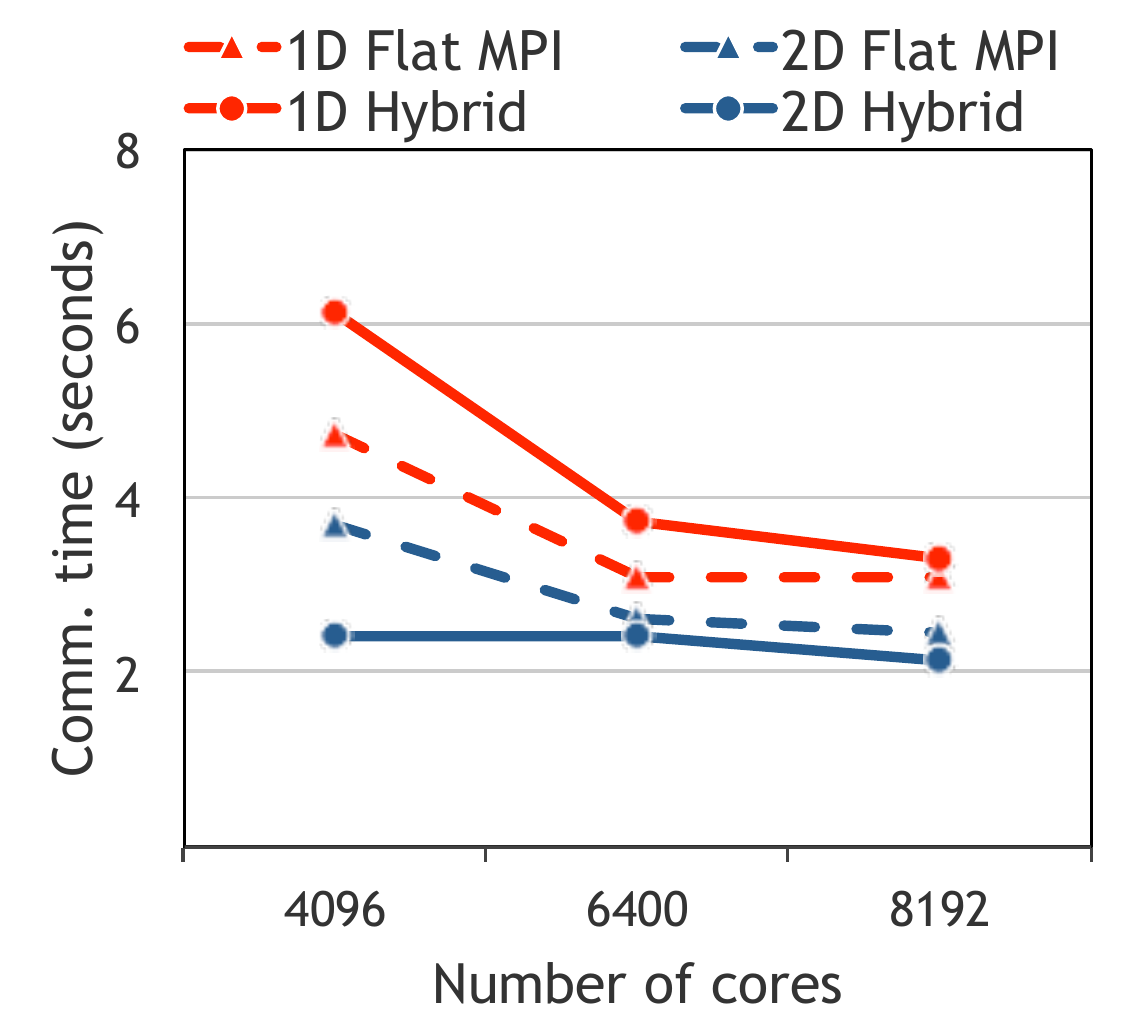}\label{fig:franklin-comm-scale32}}
\caption{BFS inter-node MPI communication time (in seconds) on Franklin for Graph 500 R-MAT graphs.}
\label{fig:franklin-commtime}
\end{figure*}

Figure~\ref{fig:franklin-strongscaling} shows `strong scaling' of our algorithms' performance (higher is better) on Franklin. 
We see that the flat 1D algorithms are about $1.5 - 1.8\times$ faster than the 2D algorithms on this architecture. The 1D hybrid algorithm, albeit slower than 
the flat 1D algorithm for smaller concurrencies, starts to perform significantly faster in larger concurrencies. We attribute this effect partially to bisection bandwidth saturation
and partially to the saturation of the network interface card when using more cores (hence more outstanding communication requests) per node. The 2D hybrid algorithm,
tends to outperform the flat 2D algorithm but can not cope with the 1D algorithms on this architecture as it spends significantly more time in computation due to
relatively larger cache working sizes, as captured by our model in Section~\ref{sec:anal}.

The communication costs, however, tell a different story about the relative competitiveness of our algorithms, Figure~\ref{fig:franklin-commtime} shows strong scaling 
of the communication time of our algorithms (lower is better). The communication times also include waiting at synchronization barriers. 
2D algorithms consistently spend less time (30-60\% for scale 32) in communication, compared to their relative 
1D algorithms. This is also expected by our analysis, as smaller number of participating processors in collective operations tend to result in faster communication times, 
with the same amount of data. The hybrid 1D algorithm catches up with the flat 2D algorithm on large concurrencies with the smaller (scale 29) dataset, even though it lags 
behind the hybrid 2D algorithm.

\begin{figure*}
\centering
\subfigure[$n=2^{30}$, $m=2^{34}$]{\includegraphics[width=0.47\textwidth]{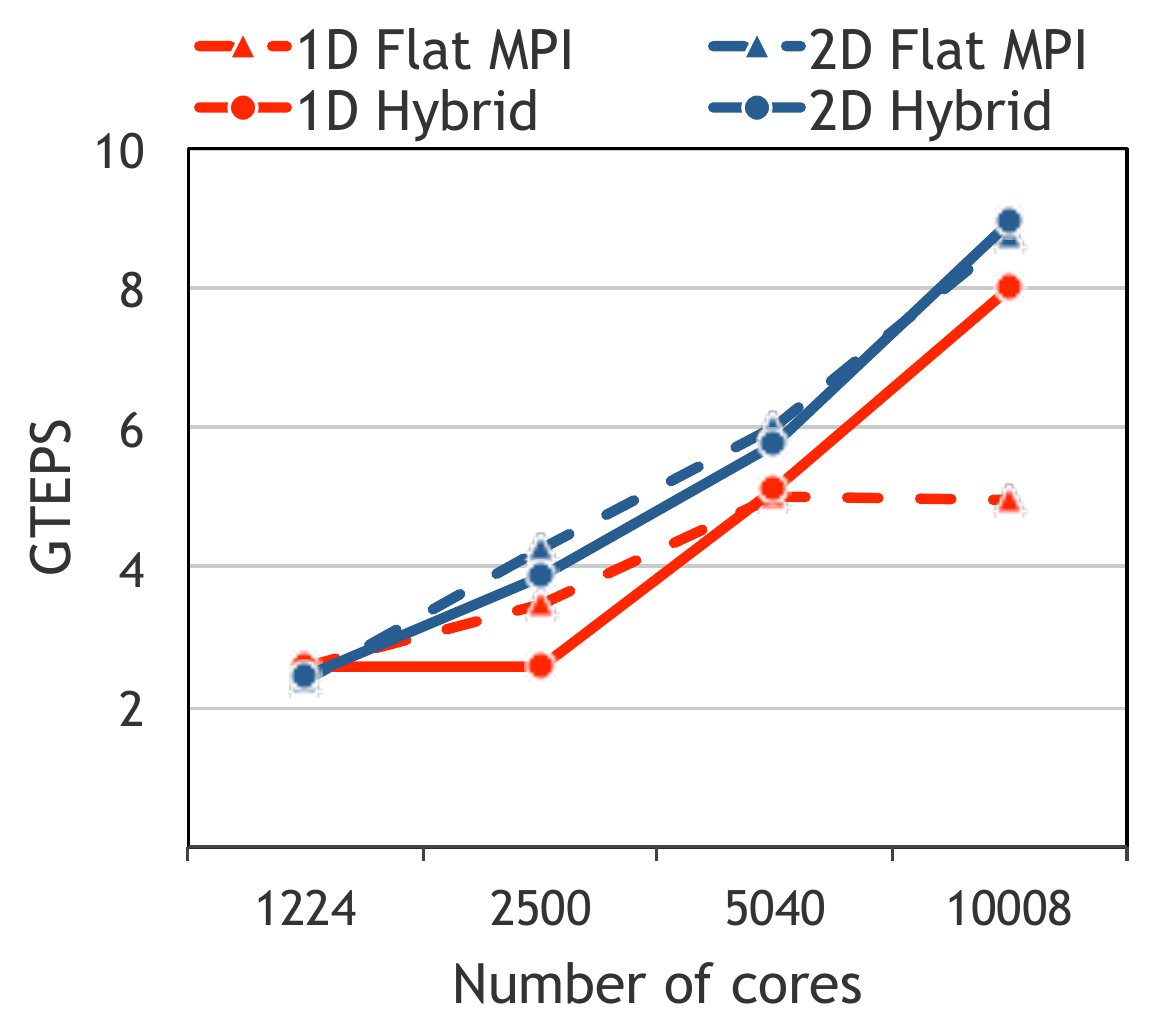}\label{fig:hopper-teps-scale30}}
\subfigure[$n=2^{32}$, $m=2^{36}$]{\includegraphics[width=0.47\textwidth]{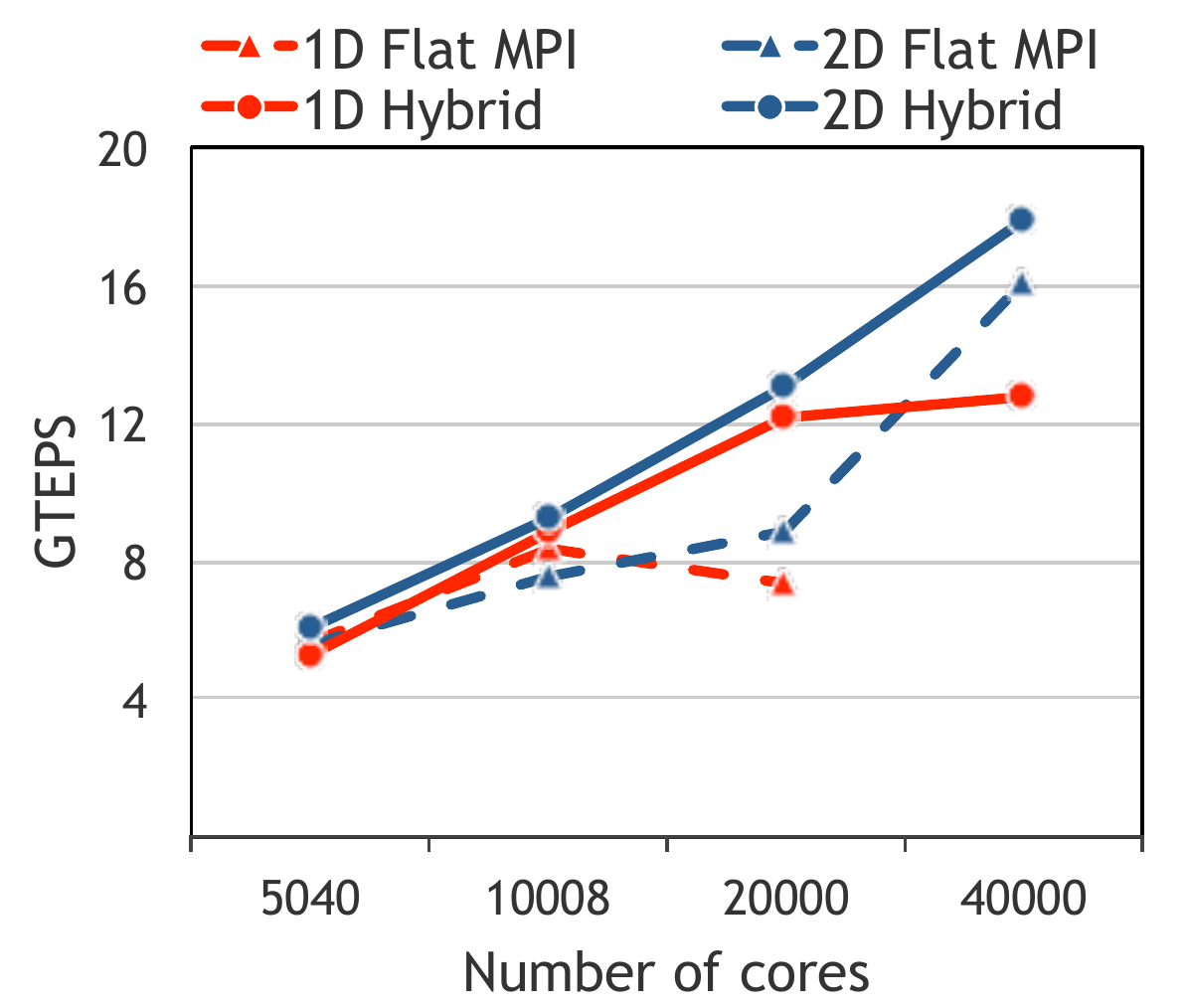}\label{fig:hopper-teps-scale32}}
\caption{BFS `strong scaling' results on Hopper for Graph 500 R-MAT graphs: Performance rate achieved (in GTEPS) on increasing the number of processors.}
\label{fig:hopper-strongscaling}
\end{figure*}

\begin{figure*}
\centering
\subfigure[$n=2^{30}$, $m=2^{34}$]{\includegraphics[width=0.47\textwidth]{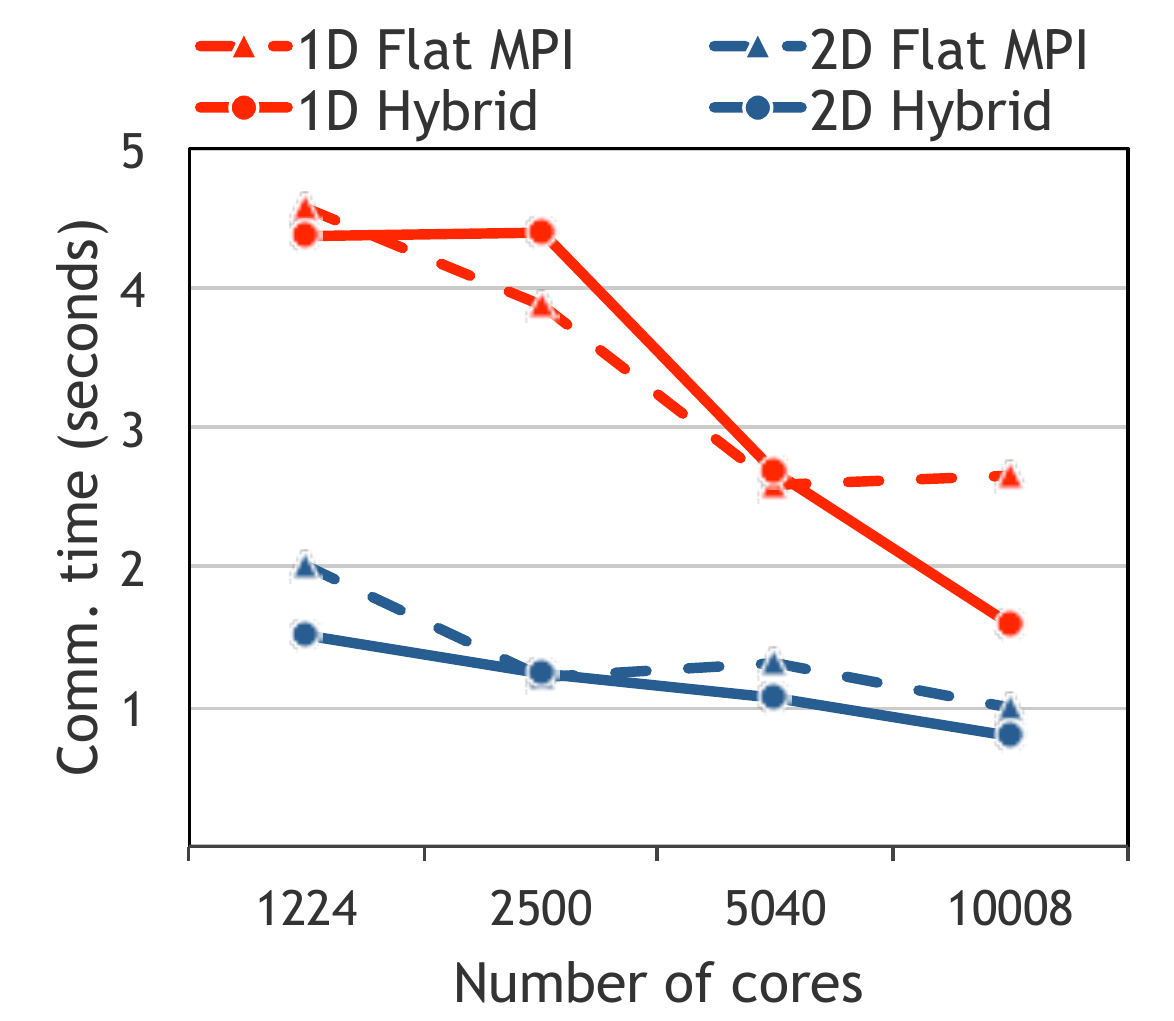}\label{fig:hopper-comm-scale30}}
\subfigure[$n=2^{32}$, $m=2^{36}$]{\includegraphics[width=0.47\textwidth]{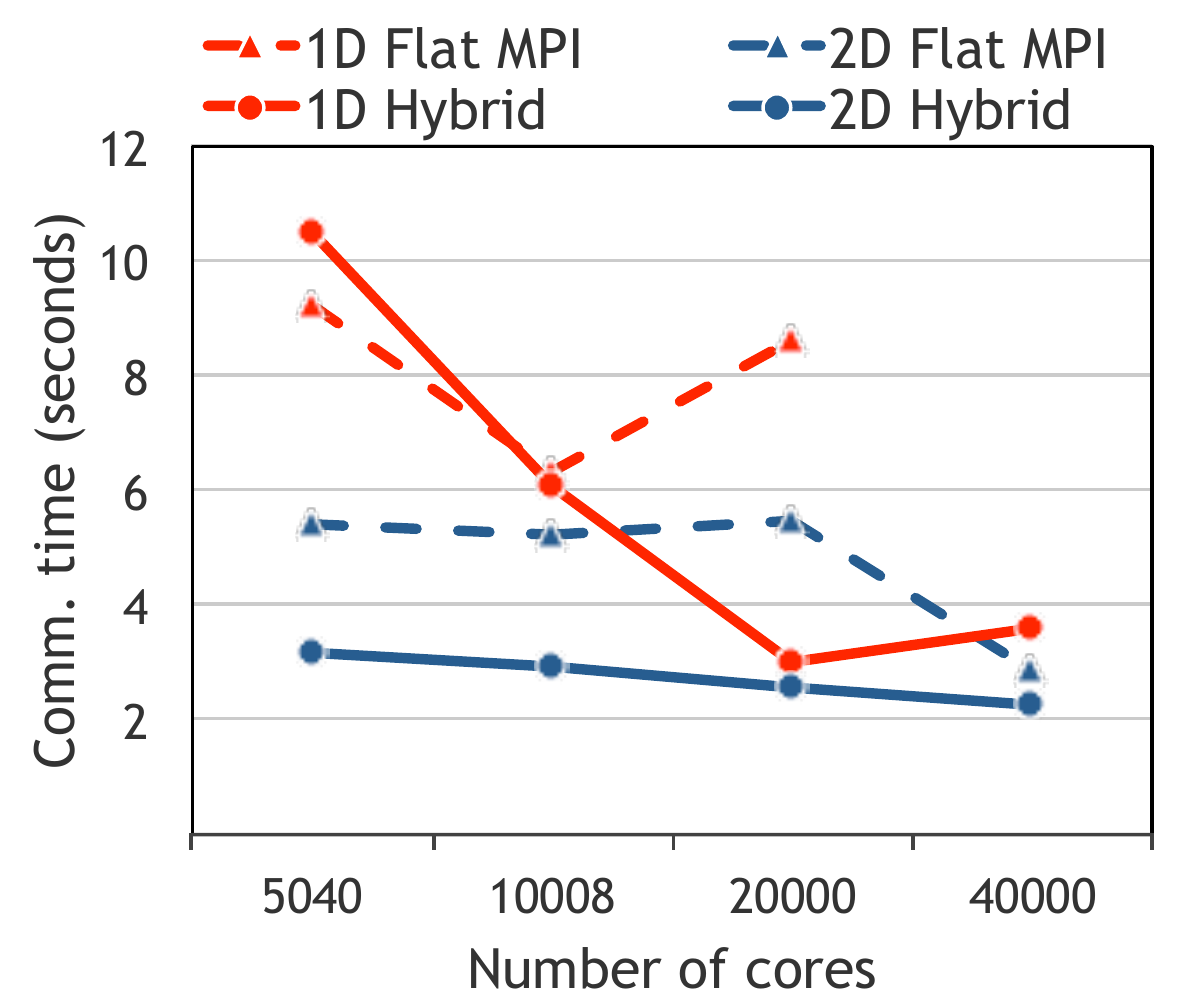}\label{fig:hopper-comm-scale32}}
\caption{BFS inter-node MPI communication time (in seconds) on Hopper for Graph 500 R-MAT graphs.}
\label{fig:hopper-commtime}
\end{figure*}

Strong scaling results on Hopper are shown in Figure~\ref{fig:hopper-strongscaling}. By contrast to Franklin results, the 2D algorithms score higher
than their 1D counterparts. The more sophisticated Magny-Cours chips of Hopper are clearly faster in integer calculations, while the overall
bisection bandwidth has not kept pace. The relative communication times of the algorithms are shown in Figure~\ref{fig:hopper-commtime}. 
We did not run the flat 1D algorithm on 40K cores as the communication times already started to increase when going from 10K to 20K cores, consuming
more than 90\% of the overall execution time. By contrast, the percentage of time spent in communication for the 2D hybrid algorithm was less than 50\% on 20K cores.

\begin{figure*}[ht]
\centering
\subfigure[Mean search time]{\includegraphics[width=0.44\textwidth]{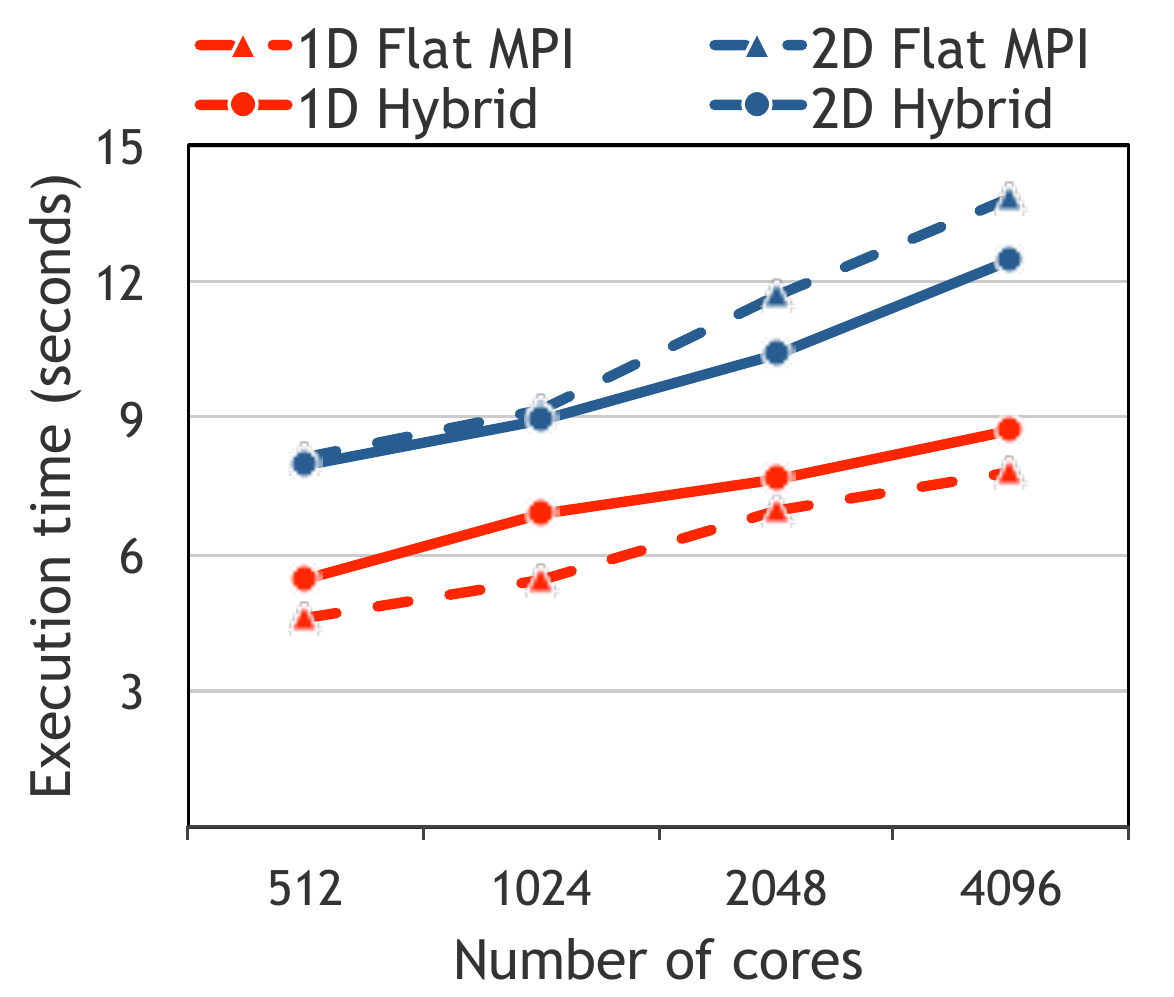}\label{fig:franklin-teps-weakscaling}}
\subfigure[Communication time]{\includegraphics[width=0.44\textwidth]{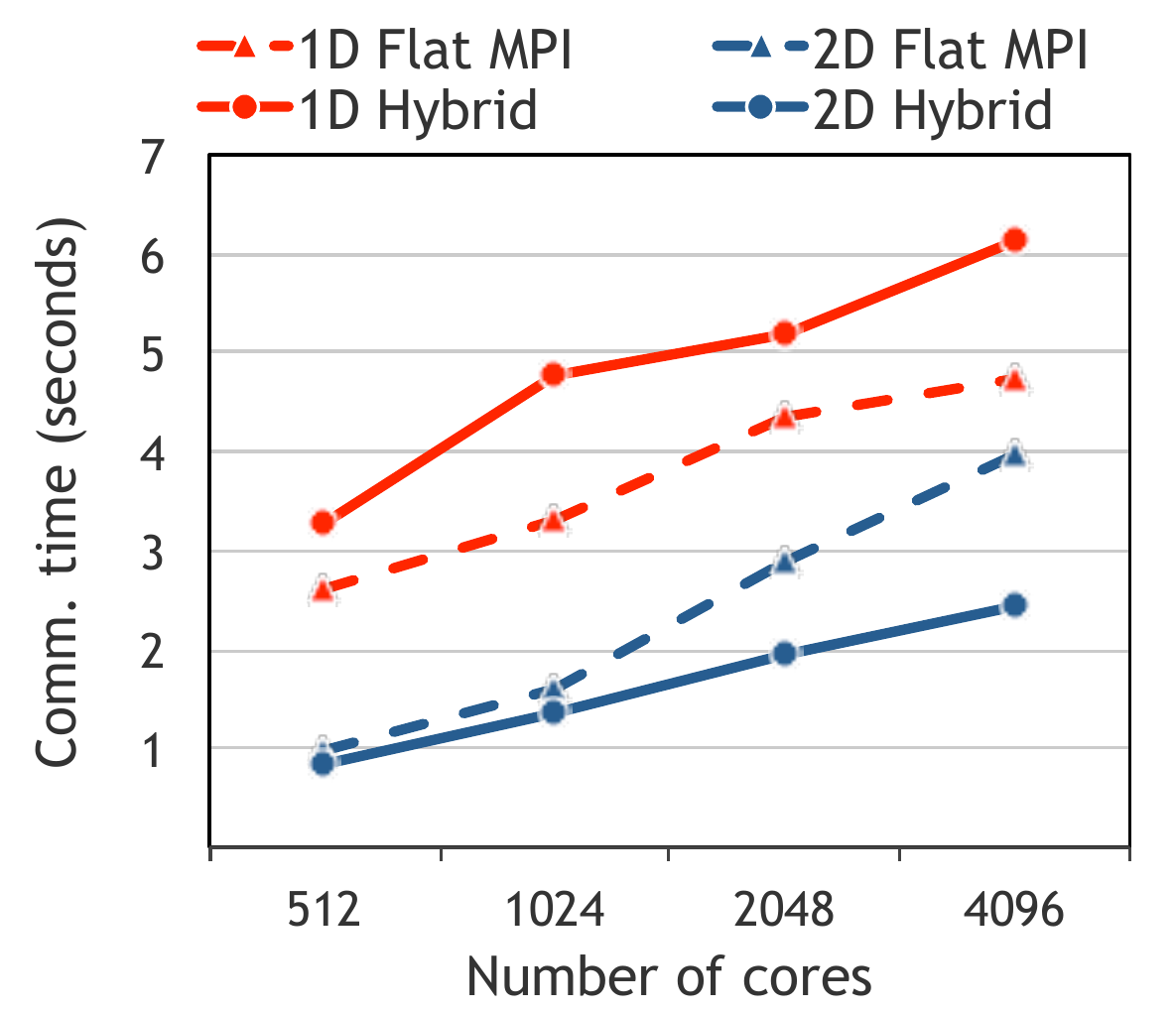}\label{fig:franklin-commtime-weakscaling}}
\caption{BFS `weak scaling' results on Franklin for Graph 500 R-MAT graphs: Mean search time (left, in seconds) and MPI communication time (right, seconds) on 
fixed problem size per core (each core has $\approx 17$M edges) For both mean search time and communication, lower is better.}
\label{fig:franklin-weakscaling}
\end{figure*}

The weak scaling results on Franklin are shown in Figure~\ref{fig:franklin-weakscaling} where we fix the edges per processor to a constant value.
To be consistent with the literature, we present weak scaling results in terms of the time it takes to complete the BFS iterations, with ideal curve being a flat line. 
Interestingly, in this regime, the flat 1D algorithm performs better than the hybrid 1D algorithm, both in terms of overall performance and
communication costs. The 2D algorithms, although performing much less communication than their 1D counterparts, come later in terms of 
overall performance on this architecture, due to their higher computation overheads.

\begin{figure*}[ht]
\centering
\subfigure[$p=1024$]{\includegraphics[width=0.44\textwidth]{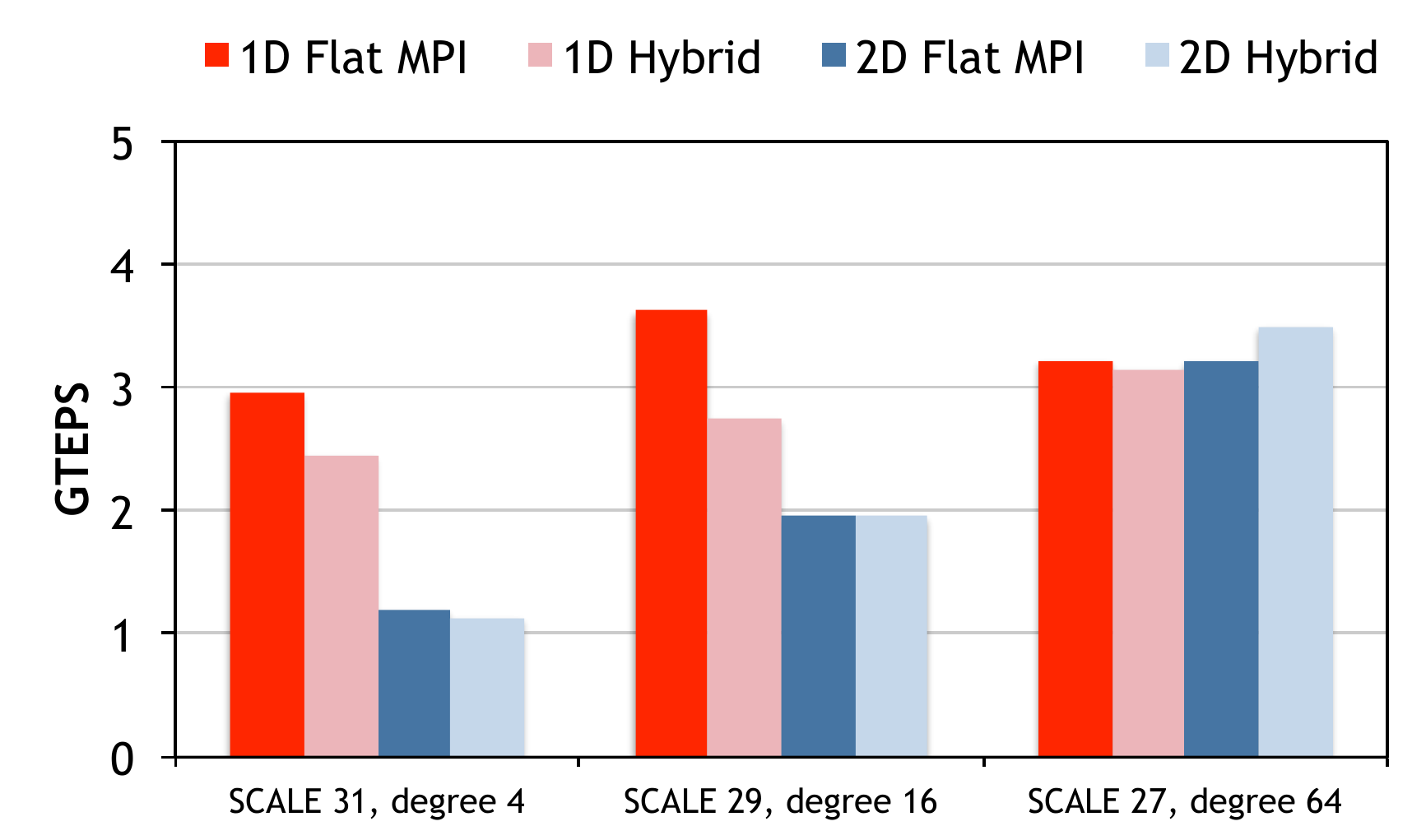}\label{fig:franklin-degreecomp-p1024}}
\subfigure[$p=4096$]{\includegraphics[width=0.44\textwidth]{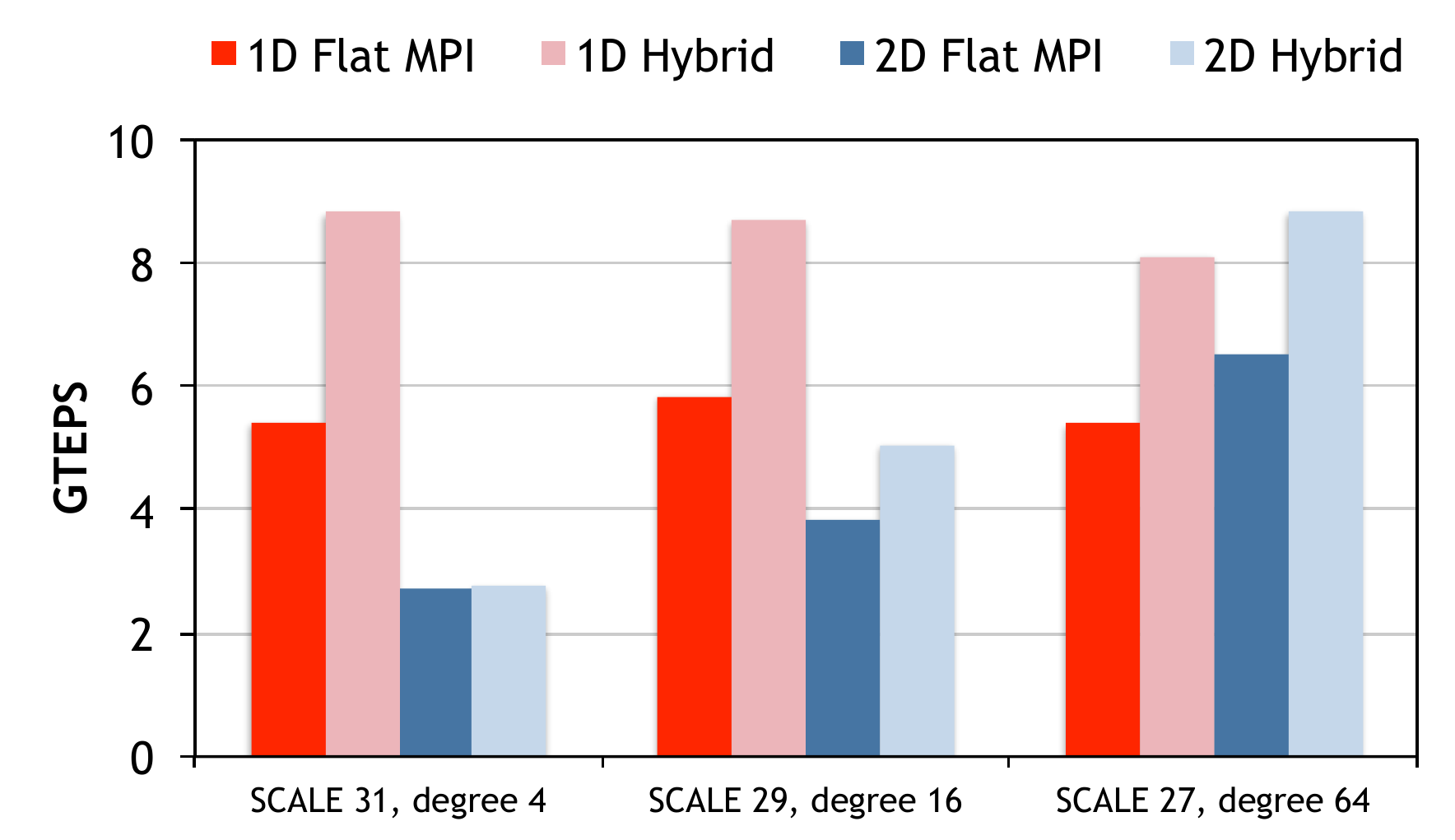}\label{fig:franklin-degreecomp-p4096}}
\caption{BFS GTEPS performance rate achieved on varying the average graph degree, for R-MAT graphs and two different parallel concurrencies.}
\label{fig:franklin-degreecomp}
\end{figure*}

Figure~\ref{fig:franklin-degreecomp} shows the sensitivity of our algorithms to varying graph densities. In this experiment, we kept the 
number of edges per processor constant by varying the number of vertices as the average degree varies. One important finding of this experiment
is that the flat 2D algorithm beats the flat 1D algorithm (for the first time) with relatively denser (average degree 64) graphs. The trend is 
obvious in that the performance margin between the 1D algorithm and the 2D algorithm increases in favor of the 1D algorithm as the graph gets
sparser. The empirical data supports our analysis in Section~\ref{sec:anal}, which stated that the 2D algorithm performance was limited by
the local memory accesses to its relatively larger vectors. For a fixed number of edges, the matrix dimensions (hence the length of
intermediate vectors) shrink as the graph gets denser, partially nullifying the cost of local cache misses.    

\begin{figure}
\centering
\includegraphics[scale=0.6]{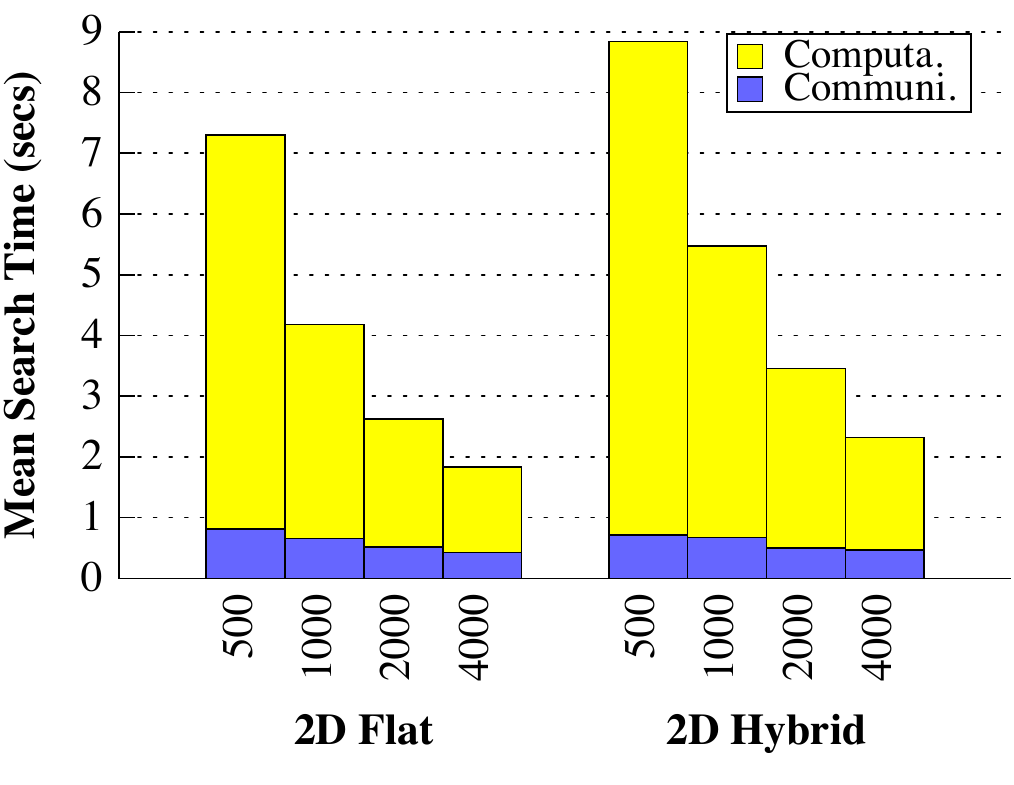}
\vsb
\caption{Running times of the 2D algorithms on the uk-union data set on Hopper (lower is better). The numbers on the x-axis is the number of cores.
The running times translate into a maximum of 3 GTEPS performance, achieving a $4\times$ speedup when going from 500 to 4000 cores}
\label{fig:ukunion}
\end{figure}

We show the performance of our 2D algorithms on the real uk-union data set in Figure~\ref{fig:ukunion}. We see that communication takes a very small
fraction of the overall execution time, even on 4K cores. This is a notable result because the uk-union dataset has a relatively high-diameter and
the BFS takes approximately 140 iterations to complete. Since communication is not the most important factor, the hybrid algorithm is slower than flat MPI, 
as it has more intra-node parallelization overheads.

To compare our approaches with prior and reference distributed memory implementations, we experimented with the Parallel Boost Graph Library's (PBGL) BFS implementation~\cite{GL05} (Version 1.45 of the Boost library) and the reference MPI implementation (Version 2.1) of the Graph 500 benchmark~\cite{graph500web}.  On Franklin, our Flat 1D code is $2.72\times$, $3.43\times$, and $4.13\times$ faster than the non-replicated reference MPI code on 512, 1024, and 2048 cores, respectively. 

Since PBGL failed to compile on the Cray machines, we ran comparison tests on Carver, 
an IBM iDataPlex system with 400 compute nodes, each node having two quad-core Intel Nehalem processors. PBGL suffers from severe memory bottlenecks
in graph construction that hinder scalable creation of large graphs. Our results are summarized in Table~\ref{tab:compare} for graphs that ran to completion. We are up to $16\times$ faster than PBGL even on these small problem instances.

Extensive experimentation reveals that our single-node multithreaded BFS version (i.e., without the inter-node communication steps in Algorithm~\ref{alg:distBFS}) is also extremely fast. The source code for some of the best x86 multicore implementations (based on absolute performance) in recent literature~\cite{APPB10,LS10} are not publicly available, and implementing these routines is non-trivial. More importantly, there is no theoretical analysis or empirical evidence to suggest these approaches would outperform ours. We compare the Nehalem-EP performance results reported in the work by Agarwal et al.~\cite{APPB10} with the performance on a single node of the Carver system (also Nehalem-EP), and notice that for R-MAT graphs with average degree 16 and 32 million vertices, our approach is nearly $1.30\times$ faster. Our approach is also faster than BFS results reported by Leiserson et al.~\cite{LS10} on the \texttt{KKt\_power}, \texttt{Freescale1}, and \texttt{Cage14} test instances, by up to $1.47\times$. 

\begin{table}[t]
\vsb
\caption{Performance comparison with PBGL on Carver. The reported numbers are in MTEPS for R-MAT graphs
with the same parameters as before. In all cases, the graphs are undirected and edges are permuted for load balance.}
\begin{center}
\begin{tabular}{c|c|c|c}
\multirow{2}{*}{Core count} 	& \multirow{2}{*}{Code} & \multicolumn{2}{c}{Problem Size} 		\\ \cline{3-4}
				&  			& Scale 22 		& Scale 24				\\ 
	\hline
\multirow{2}{*}{128}		& PBGL 		& $25.9$ 		& $39.4$ 			\\
					& Flat 2D 		& $266.5$  	& $567.4$  			\\
\hline
\multirow{2}{*}{256}		& PBGL 		& $22.4$ 		& $37.5$  			\\
					& Flat 2D 		& $349.8$ 	& $603.6$  			
\end{tabular}
\label{tab:compare}
\end{center}
\end{table}

\section{Conclusions and Future Work}

In this paper, we present a design-space exploration of distributed-memory parallel BFS, discussing two fast ``hybrid-parallel'' approaches for large-scale graphs. Our experimental study encompasses performance analysis on several large-scale synthetic random graphs that are also used in the recently announced Graph 500 benchmark. The absolute performance numbers we achieve on the large-scale parallel systems Hopper and Franklin at NERSC are significantly higher than prior work. The performance results, coupled with our analysis of communication and memory access costs of the two algorithms, challenges conventional wisdom that fine-grained communication is inherent in parallel graph algorithms and necessary for achieving high performance~\cite{LGH07}. 

We list below optimizations that we intend to explore in future work, and some open questions related to design of distributed-memory graph algorithms.





\noindent \textbf{Exploiting symmetry in undirected graphs.} If the graph is undirected, then one can save 50\% space by storing only the upper (or lower) triangle of the sparse adjacency matrix, effectively doubling the size of the maximum problem that can be solved in-memory on a particular system. The algorithmic modifications needed to save a comparable amount in communication costs for BFS iterations is not well-studied.  

\noindent \textbf{Exploring alternate programming models.} Partitioned global address space (PGAS) languages can potentially simplify expression of graph algorithms,  as inter-processor communication is implicit. In future work, we will investigate whether our two new BFS approaches are amenable to expression using PGAS languages, and whether they can deliver comparable performance. 

\noindent \textbf{Reducing inter-processor communication volume with graph partitioning.} 
An alternative to randomization of vertex identifiers is to use hypergraph partitioning software to reduce communication. Although hypergraphs are
capable of accurately modeling the communication costs of sparse matrix-dense vector multiplication, SpMSV has not 
been studied yet, which is potentially harder since the sparsity pattern of the frontier matrix changes over BFS iterations.

\noindent \textbf{Interprocessor collective communication optimization.} We conclude that even after alleviating the communication costs, the performance of distributed-memory parallel BFS is heavily dependent on the inter-processor collective communication routines All-to-all and Allgather. Understanding the bottlenecks in these routines at high process concurrencies, and designing network topology-aware collective algorithms is an interesting avenue for future research.

\section*{Acknowledgments}
Discussions with John R.\,Gilbert, Steve Reinhardt, and Adam Lugowski greatly improved our understanding of casting BFS iterations into sparse linear algebra.
John Shalf and Nick Wright provided generous technical and moral support during the project.

\bibliographystyle{plain} 
\bibliography{parallel,bfs}
\end{document}